\documentclass[aps,11pt,prc,superscriptaddress,nofootinbib]{revtex4}

\usepackage[usenames]{color}
\usepackage[dvips]{graphicx}
\usepackage{amsmath}
\usepackage{amsfonts}
\usepackage{amssymb}
\usepackage{mathrsfs}
\usepackage{bm}
\usepackage{verbatim}
\usepackage{cancel}

\newcommand{\beq}{\begin{equation}}
\newcommand{\eeq}{\end{equation}}
\newcommand{\bea}{\begin{eqnarray}}
\newcommand{\eea}{\end{eqnarray}}

\newcommand{\mlo}{M_{\text{lo}}}
\newcommand{\mhi}{M_{\text{hi}}}
\newcommand{\chn}[3]{{{}^{#1}{#2}_{#3}}}
\newcommand{\cs}[2]{\chn{#1}{S}{#2}}
\newcommand{\cp}[2]{\chn{#1}{P}{#2}}
\newcommand{\cd}[2]{\chn{#1}{D}{#2}}
\newcommand{\cf}[2]{\chn{#1}{F}{#2}}
\newcommand{\csd}{{\cs{3}{1}-\cd{3}{1}}}
\newcommand{\cpf}{{\cp{3}{2}-\cf{3}{2}}}
\newcommand{\mnda}{\overline{\text{NDA}}}

\begin{document}

\title{Renormalizing Chiral Nuclear Forces: Triplet Channels}
\author{Bingwei Long}
\email{bingwei@jlab.org}
\affiliation{Excited Baryon Analysis Center (EBAC), Jefferson Laboratory, 12000 Jefferson
Avenue, Newport News, Virginia 23606, USA}
\author{C.-J. Yang}
\email{cjyang@email.arizona.edu}
\affiliation{Department of Physics, University of Arizona, Tucson, Arizona 85721, USA}
\affiliation{Institute of Nuclear and Particle Physics, Ohio University, Athens, Ohio
45701, USA}

\date{\today}

\preprint{INT-PUB-11-038}
\preprint{JLAB-THY-11-1464}

\begin{abstract}
We discuss the subleading contact interactions, or counterterms, of the triplet channels of nucleon-nucleon scattering in the framework of chiral effective field theory, with $S$ and $P$ waves as the examples. The triplet channels are special in that they allow the singular attraction of one-pion exchange to modify Weinberg's original power counting (WPC) scheme. With renormalization group invariance as the constraint, our power counting for the triplet channels can be summarized as a modified version of naive dimensional analysis in which, when compared with WPC, all of the counterterms in a given partial wave (leading or subleading) are enhanced by the same amount. More specifically, this means that WPC needs no modification in $\csd$ and $\cp{3}{1}$, whereas a two-order enhancement is necessary in both $\cp{3}{0}$ and $\cpf$. 
\end{abstract}

\maketitle

\section{Introduction\label{sec:intro}}

In applying the idea of chiral effective field theory (EFT) to nuclear physics, a great deal of effort has been devoted to implementing Weinberg's original prescription~\cite{Weinberg:1990-1991} for the problems of few-nucleon systems, with the two-nucleon system as the starting point~\cite{Ordonez:1993-1995, Epelbaum:1998ka-1999dj, Epelbaum:2004fk, Entem:2001cg-2002sf, PavonValderrama:2005uj, YangOhio, saopaulo} (for more general reviews, see Refs.~\cite{vanKolck:1999mw, Beane:2000fx, Bedaque:2002mn, Epelbaum:2005pn, Epelbaum:2008ga, Machleidt:2010kb}). While the power counting of pion exchange diagrams follows the paradigm of chiral perturbation theory (ChPT), i.e., chiral EFT in the single-nucleon sector, estimating the size of $NN$ contact interactions often requires assumptions beyond chiral symmetry.

Assumed in Weinberg's power counting (WPC) is what we refer to as naive dimensional analysis (NDA): each derivative on or each power of pion mass dependence of the Lagrangian terms is always suppressed by the underlying scale of chiral EFT, $M_{\text{hi}} \sim m_\sigma$, where $m_\sigma$ is the mass of the $\sigma$ meson. While plausible, this assumption was questioned in a number of works~\cite{Kaplan:1996xu, Beane:2000wh, Beane:2001bc, Nogga:2005hy, Birse:2005um, Birse:2009my, Valderrama:2009ei, Valderrama:2011mv, Long:2011qx}, and was shown to be only partially correct from the perspective of renormalization group (RG) invariance; WPC does not have enough $NN$ contact interactions, or counterterms, to renormalize the nucleon-nucleon ($NN$) scattering amplitudes at a given order. One mechanism to spoil the renormalizability of WPC is the singular---diverging at least as fast as $1/r^2$---attraction of the tensor force of one-pion exchange (OPE): $-1/r^3$ at $r \to 0$. In $S$ and $P$ waves, the triplet channels subject to this singular attraction include one uncoupled, $\cp{3}{0}$, and two coupled, $\csd$ and $\cpf$. In Ref.~\cite{Long:2011qx}, we used $\cp{3}{0}$ to investigate the modification to WPC under the guidance of RG invariance. In this paper, we report a complete study of the triplet channels, in which we continue our efforts to modify WPC at the subleading orders, in a generalization of Ref.~\cite{Long:2007vp}. Interestingly, we reach conclusions that differ in some aspects from a parallel investigation in Refs.~\cite{Valderrama:2009ei, Valderrama:2011mv}.

Except for the attempts to treat OPE as perturbation~\cite{Kaplan:1996xu, Kaplan:1998tg, Fleming:1999ee, Beane:2008bt}, it is well accepted that in $S$ and $P$ waves the leading order (LO) amplitude requires the full iteration of OPE~\cite{Birse:2005um, Nogga:2005hy}. Even though the nonperturbative unitarity requires any nonperturbative, nonrelativistic $T$-matrix to scale as $Q^{-1}$, we \emph{choose} to label the LO as $\mathcal{O}(1)$ so that one does not need to change the standard ChPT notation for power counting pion exchange diagrams. What is more consequential is that we denote different orders of the EFT expansion by its relative correction to the LO, i.e., the next-to-leading order (NLO) by $\mathcal{O}(Q/M_{\text{hi}})$ or $\mathcal{O}(Q)$ for short, and next-to-next-to-leading order (NNLO) by $\mathcal{O}(Q^{2}/M_{\text{hi}}^{2})$ or $\mathcal{O}(Q^2)$, and so on.\footnote{Here NLO and NNLO are defined differently from a more conventional notation~\cite{Epelbaum:1998ka-1999dj, Entem:2001cg-2002sf}, where NLO is $\mathcal{O}(Q^{2})$ and NNLO is $\mathcal{O}(Q^{3})$.} RG invariance, or, more specifically, invariance of the amplitude with respect to the ultraviolet (UV) momentum cutoff inherent in the Lippmann-Schwinger equation, demands as well one counterterm to be fully iterated in the partial waves where the tensor OPE is attractive~\cite{Nogga:2005hy}. In $P$ waves, these singular attractive channels include $\cp{3}{0}$ and $\cpf$. However, WPC considers the counterterms in these channels to be subleading, for they are second-order polynomials in momenta. Stated differently, the leading counterterms in $\cp{3}{0}$ and $\cpf$ are underestimated in WPC, and RG invariance requires them to be enhanced by $\mathcal{O}(M_{\text{hi}}^2/M_{\text{lo}}^2)$. Here, $M_{\text{lo}}$ refers to a cluster of infrared mass scales that include the pion decay constant $f_\pi \simeq 92$ MeV, the pion mass $m_\pi \simeq 140$ MeV, and certain combinations of them.

Although not one of the centerpieces of WPC, the indiscriminate, full iteration of different order potentials as a whole has been the standard practice in its implementations. However, the ordering of the potentials according to their matrix elements for small momenta is not always valid in the nonperturbative treatment in which the intermediate states could reach $\mhi$, where the higher-order potentials usually have larger matrix elements. In the nonperturbative setup, it seems to take intricate cancellations for higher-order potentials to eventually contribute less to the low-energy, on-shell amplitude. But it is far beyond the scope of our paper to decide whether or under what conditions these cancellations will happen. We refer readers to Refs.~\cite{Epelbaum:2009sd, Entem:2007jg} for discussion regarding renormalization and power counting in the nonperturbative treatment.

To minimize the interference between lower- and higher-order potentials in the UV region, we choose the natural way to go beyond the LO, that is, to treat the subleading interactions as perturbations. Now that the potentials from different orders are no longer on an equal footing, it is, as we will see, much easier to separate in the UV region the contributions of higher-order potentials from those of lower-order ones. If the subleading interactions are too strong to be perturbative, they are simply not subleading in a \textit{bona fide} EFT. Reference~\cite{Long:2007vp} explained the perturbative formalism with a toy model: $-1/r^{2}$ as LO and $\pm 1/r^{4}$ as $\mathcal{O}(Q^2)$ long-range potentials. More importantly, the general lesson drawn from the study of Ref.~\cite{Long:2007vp}, referred to in the paper as modified NDA (\,$\mnda$\,), is that in the case of the LO long-range potential being singular and attractive, the subleading counterterms (SCTs) are enhanced relative to NDA by the same amount as the LO counterterms; as the long-range force gets an $\mathcal{O}(Q^2)$ correction, so do the contact operators that have two more derivatives than the LO counterpart.

The validity of $\mnda$ is confirmed by renormalization of uncoupled $\cp{3}{0}$ up to $\mathcal{O}(Q^3)$~\cite{Valderrama:2011mv, Long:2011qx}, which shows that the leading long-range potential being $-1/r^2$ is not essential to the applicability of $\mnda$. Now $\csd$ and $\cpf$ pose interesting questions as to the extension of $\mnda$ to the coupled-channel problems.
Take $\csd$ as an example. The LO counterterm is a constant, $C_\cs{3}{1}$, and, according to $\mnda$, two second-derivative terms will turn up at $\mathcal{O}(Q^2)$. Therefore, $\mnda$ suggests a total of three counterterms up to $\mathcal{O}(Q^3)$ in each of $\csd$ and $\cpf$, which will be discussed in more details in Sec.~\ref{sec:cc}. However, using a coordinate space setup, Refs.~\cite{Valderrama:2009ei, Valderrama:2011mv} concluded that there must be six counterterms in $\csd$ or $\cpf$ up to $\mathcal{O}(Q^3)$. If this proliferation of counterterms in the coupled channels is true, then the predictive power of nuclear EFT is further weakened.

We carry out in the paper a momentum-space calculation to verify the power counting based on $\mnda$. In particular, we are interested to see which can be confirmed in the coupled channels: the proliferation of six counterterms or three counterterms inferred from $\mnda$.
Our principle of establishing power counting is summarized as follows:
\begin{itemize}
  \item[(i)] The size of pion exchanges is decided by the non-analytic part of the corresponding Feynman diagram, which is rightly captured by WPC.
  \item[(ii)] We promote counterterms over WPC only if RG invariance requires it. That is, when a counterterm is not needed for renormalization, its counting will follow NDA.
\end{itemize}
The rationale for the second point is that RG analysis in terms of the floating momentum cutoff of the $NN$ intermediate states  touches upon only (nonrelativistic) nucleon momenta and it does not ``know'' anything about the contributions of heavy mesons that are integrated out in the first place~\cite{bira-private}.

A study of power counting of chiral $NN$ forces is not complete without the singlet channels. But the drastically different short-range behavior of OPE in the singlet ($1/r$) and triplet ($1/r^3$) channels signals different structures of counterterms. Therefore, we leave the singlet channels to a further study~\cite{BwLCJY-singlet}.

We also leave out $D$ and higher waves in our analysis, except for $\cd{3}{1}$ and $\cf{3}{2}$ which are, respectively, coupled to $\cs{3}{1}$ and $\cp{3}{2}$. The reasons are as follows. First, the conceptual issues concerning renormalization and power counting can be well illustrated by $S$ and $P$ waves. Second, it is debatable whether or to what extent OPE is perturbative in $D$ waves~\cite{Kaiser:1997mw, Kaiser:1998wa, Entem:2001cg-2002sf, Birse:2005um}, and answering this question is beyond the scope of our paper. 

After briefly reviewing the LO and establishing our notation in Sec.~\ref{sec:lo}, we will establish in Sec.~\ref{sec:subcon} the SCTs in $\cp{3}{1}$, $\cp{3}{0}$, $\csd$ and $\cpf$ by examining the cutoff dependence of the subleading amplitudes. Finally, we offer a discussion and a conclusion in Sec.~\ref{sec:conclusion}.

\section{Leading Order\label{sec:lo}}

OPE is the leading long-range $NN$ interaction,
\begin{equation}
V_{L}^{(0)}(\vec{q}\,)=V_{1\pi}(\vec{q}\,)\equiv -\frac{g_{A}^{2}}{4f_{\pi}^{2}}\,\bm{\tau}_{1}\bm{\cdot}\bm{\tau}_{2}\,\frac{\vec{\sigma}_{1}\cdot\vec{q}\,\vec{\sigma}_{2}\cdot\vec{q}}{{\vec{q}}\,^{2}+m_{\pi}^{2}}\,,
\end{equation}
where $\vec{q}\equiv \vec{p}\,'-\vec{p}$ is the difference between the outgoing ($\vec{p}\,'$) and the incoming ($\vec{p}\,$) momenta in the center-of-mass frame, the axial vector coupling constant $g_{A}=1.29$, the pion decay constant $f_{\pi}=92.4$ MeV, and the pion mass $m_{\pi}=138$ MeV. Its coordinate space version will be useful in the discussion,
\begin{equation}
V_{1\pi}(\vec{r}\,) = \lambda_\pi \bm{\tau}_{1}\bm{\cdot}\bm{\tau}_{2}\, \left[T(r) S_{12} + Y(r) \vec{\sigma}_1\cdot\vec{\sigma}_2 \right] \, ,
\end{equation} 
where
\begin{align}
\lambda_\pi &= \frac{m_\pi^3}{12\pi}\left(\frac{g_{A}^{2}}{4f_{\pi}^{2}}\right) \, , \\
T(r) &= \frac{e^{-m_\pi r}}{m_\pi r} \left[1 + \frac{3}{m_\pi r} + \frac{3}{(m_\pi r)^2} \right] \, , \\
Y(r) &= \frac{e^{-m_\pi r}}{m_\pi r} \, ,
\end{align}
and
\begin{equation}
S_{12} = 3(\vec{\sigma}_1\cdot\hat{r})(\vec{\sigma}_2\cdot\hat{r}) - \vec{\sigma}_1\cdot\vec{\sigma}_2 \, .
\end{equation}
The tensor force $T(r)$ has an inverse cubic short-range core, $1/r^3$, but it contributes to only triplet channels. 

In the lower partial waves, the LO amplitude $T^{(0)}$ is obtained by the full iteration of OPE and necessary counterterms, through solving the Lippmann-Schwinger equation. For the coupled channel with total angular momentum $j$, the off-shell Lippmann-Schwinger equation reads
\begin{equation}
T^{(0)}_{l' l}(p', p; k) = V^{(0)}_{l' l}(p', p) +  \frac{2}{\pi} m_N \sum_{l''} \int^\Lambda d\kappa\, \kappa^2\, V^{(0)}_{l' l''}(p', \kappa) \frac{T^{(0)}_{l'' l}(\kappa, p; k)}{k^2 - \kappa^2 + i\epsilon} \, ,\label{eqn:LSE0}
\end{equation} 
with $l$, $l'$, and $l''$ running over $j-1$ and $j+1$, $k$ as the center-of-mass momentum, and $\Lambda$ as the momentum cutoff. Extension to the uncoupled channels is straightforward. The product of the Schr\"odinger propagator and the integral measure scales as $m_N Q$. Since OPE scales, more or less casually, as $(m_N \mlo)^{-1}$, with $\mlo$ a certain combination of $f_\pi^2$ and $m_N$, OPE must become nonperturbative when $Q \sim \mlo$.

Depending on the sign of the matrix element of $S_{12}$, the OPE tensor force drives $NN$ contact interactions in very different ways~\cite{Nogga:2005hy}. This is best elucidated in the uncoupled channels. When $\langle lsj|S_{12}|lsj \rangle$ is positive, where $l$ is the orbital angular momentum, $s=1$ is the spin, and $j$ is the total angular momentum, the OPE tensor force is $\sim +1/r^3$. If one picks up the regular solution to this repulsive potential, the wave function dies off exponentially near the origin. As a consequence, the sensitivity to the UV cutoff vanishes very quickly; thus, there is no need for an extra counterterm to absorb the cutoff dependence~\cite{PavonValderrama:2005uj}. This is in agreement with WPC because, according to WPC, the first counterterm in $\cp{3}{1}$---the lowest repulsive, uncoupled triplet channel---appears at $\mathcal{O}(Q^2)$.

When $\langle lsj|S_{12}|lsj \rangle$ is negative, the OPE tensor force overpowers the kinetic energy and the centrifugal barrier, causing the $NN$ system to ``collapse''~\cite{landau}. The mathematical origin of this pathology is the simultaneous existence of two equally good solutions to the Schr\"odinger equation~\cite{PavonValderrama:2005uj}, which eventually lead to ambiguity in predicting physical observables. Or, in terms of the Lippmann-Schwinger equation, the scattering amplitude is very sensitive to the UV cutoff.

The modern-day interpretation of renormalization offers a cure to this sort of pathological potential with singular attraction near the origin: supplementing short-range interactions $V_S$ rather than naively extrapolating the long-range interaction to short distances. The model-independent treatment involves arranging $V_S$ as counterterms that run with the UV cutoff $\Lambda$ in such a way that physical observables do not depend on $\Lambda$~\cite{Birse:1998dk, Beane:2000wh, Barford:2002je, Birse:2005um}. Although this is not a complete innovation in the context of chiral EFT since $NN$ contact terms are always part of the chiral Lagrangian, the consideration of RG invariance offers \textit{a priori} insights into gauging the importance of counterterms at a given order.

In $\cp{3}{0}$ where $l > 0$, the singular attraction of OPE requires at $\mathcal{O}(1)$ a counterterm, which is, however, considered $\mathcal{O}(Q^2)$ in WPC for it is a second order polynomial in momenta~\cite{Nogga:2005hy},
\begin{equation}
\langle\, \cp{3}{0} | V_S^{(0)} |\, \cp{3}{0} \rangle = C_\cp{3}{0} p' p \, ,
\label{eqn:LOuncpldCC}
\end{equation} 
where $p'$ ($p$) is the magnitude of $\vec{p}\,'$ ($\vec{p}\,$). Now that $C_\cp{3}{0} p' p$ is at LO with OPE $\sim 4\pi/(m_N \mlo)$, $C_\cp{3}{0}$ must scale as
\begin{equation}
C_\cp{3}{0} \sim \frac{4\pi}{m_N} \frac{1}{\mlo^3} \, ,
\end{equation}
where $4\pi/m_N$ is introduced to cancel a common factor that usually concurs with loop integrals involving $NN$ intermediate states. For comparison, WPC considers $C_\cp{3}{0} p' p$ to be of the same order as the leading two-pion exchange, which in turn is counted as an $\mathcal{O}(Q^2/\mhi^2)$ correction to OPE; therefore, WPC prescribes~\cite{vanKolck:1999mw}
\begin{equation}
C^\text{WPC}_\cp{3}{0} \sim \frac{4\pi}{m_N} \frac{1}{\mlo\mhi^2} \, .
\end{equation}

In the coupled channels, it is convenient to write the tensor OPE in coordinate space as a $2\times 2$ matrix in the basis of two coupled orbital angular momentum states, $l = j \pm 1$~\cite{goldberger}. In $\cs{3}{1} - \cd{3}{1}$ and $\cp{3}{2} - \cf{3}{2}$,
\begin{align}
V_{T 1\pi}(\cs{3}{1} - \cd{3}{1}) &= \lambda_\pi
\begin{pmatrix}
0 & -6\sqrt{2} \\
-6\sqrt{2} & 6
\end{pmatrix}
T(r) \, , \label{eqn:VT1pi3s1} \\
V_{T 1\pi}(\cp{3}{2} - \cf{3}{2}) &= \frac{\lambda_\pi}{5}
\begin{pmatrix}
-2 & -8 \\
-8 & 6\sqrt{6}
\end{pmatrix}
T(r) \, . \label{eqn:VT1pi3p2}
\end{align}
There is a domain of $r$ near the origin in which $T(r)$ dominates over the centrifugal barrier and in which diagonalizing $V_{T 1\pi}$ also diagonalizes the Schr\"odinger equation.
OPE matrix elements in all of the coupled channels
share one property: there always is one attractive and one repulsive eigen subchannel. Therefore, one needs to summon at LO a short-range input to counter the singular attraction in one of the subchannels.
This accords with WPC in $\cs{3}{1} - \cd{3}{1}$ but calls for amendment in $\cp{3}{2} - \cf{3}{2}$~\cite{Nogga:2005hy}.

Although the counterterms are easily formulated as polynomials in momentum space, the above-mentioned diagonalization cannot be trivially realized therein. The counterterms in $\csd$ and $\cpf$ have the generic momentum space form
\begin{align}
\langle \csd | V_S | \csd \rangle &=
\begin{pmatrix}
C_\cs{3}{1} + D_\cs{3}{1}({p'}^2 + p^2) & E_\text{SD}\,p^2 \\
E_\text{SD}\, {p'}^2 & F_\cd{3}{1}\, {p'}^2p^2
\end{pmatrix} + \cdots \, , \label{eqn:3s1gnrVS} \\
\langle \cpf | V_S | \cpf \rangle &= p' p
\begin{pmatrix}
C_\cp{3}{2} + D_\cp{3}{2}({p'}^2 + p^2) & E_\text{PF}\,p^2 \\
E_\text{PF}\, {p'}^2 & F_\cf{3}{2}\, {p'}^2p^2
\end{pmatrix} + \cdots \, . \label{eqn:3p2gnrVS}
\end{align} 
As shown in Ref.~\cite{Nogga:2005hy}, it is not necessary to design a counterterm that exclusively acts on the attractive subchannel of $V_{T 1\pi}$ \eqref{eqn:VT1pi3s1}; the $C$ term alone
will properly renormalize all of the $T$-matrix elements.

While the $C$ term in $\csd$ does not violate WPC, the promotion of the $C$ term in $\cpf$
asks for an enhancement of $\mathcal{O}(\mhi^2/\mlo^2)$:
\begin{align}
\langle\, \csd | V_S^{(0)} |\, \csd \rangle &=
\begin{pmatrix}
C_\cs{3}{1}^{(0)} & 0 \\
0 & 0
\end{pmatrix} \, , \\
\langle\, \cpf | V_S^{(0)} |\, \cpf \rangle &= p' p
\begin{pmatrix}
C_\cp{3}{2}^{(0)} & 0 \\
0 & 0
\end{pmatrix} \, .
\label{eqn:c3p2lo}
\end{align}
Stated differently, the renormalized $C_\cs{3}{1}$ and $C_\cp{3}{2}$ scale as
\begin{equation}
C_\cs{3}{1} \sim \frac{4\pi}{m_N} \frac{1}{\mlo}\, , \qquad C_\cp{3}{2} \sim \frac{4\pi}{m_N} \frac{1}{\mlo^3} \, ,
\end{equation}
whereas WPC differs for $C_\cp{3}{2}$,
\begin{equation}
C_\cp{3}{2}^\text{WPC} \sim \frac{4\pi}{m_N} \frac{1}{\mlo\mhi^2} \, .
\end{equation}

\section{Subleading Orders\label{sec:subcon}}

\subsection{Generalities}

It has been long known that the two-pion exchanges (TPEs) with chiral index $\nu =0$ vertices (TPE0) give an $\mathcal{O}(Q^{2})$ correction to OPE, and that the TPEs with one insertion of $\nu =1$ vertices (TPE1) lead to $\mathcal{O}(Q^{3})$ long-range potentials. TPEs are not uniquely defined because different regularization schemes---such as dimensional and spectral function regularizations~\cite{Epelbaum:2003gr2003xx}---applied to two-pion-exchange Feynman diagrams may lead to different expressions. To remove the arbitrariness in the choice of regulator, one should always pair TPE expressions with a momentum polynomial.
We refer to this polynomial as the ``primordial'' counterterm for TPEs. By definition, the primordial counterterm is power counted as the same order as the corresponding TPEs, which is exactly the content of WPC.

In addition to the general rationale given at the end of Sec.~\ref{sec:intro}, the concept of primordial counterterms serves as a complementary argument against demoting counterterms in the cases where they are not needed for renormalization in the context of the Lippmann-Schwinger equation.

TPEs, as the two-pion-exchange diagrams evaluated in the plane-wave basis, are not the only sources driving contact interactions when high momentum modes are integrated out. There are two other classes of diagrams that contribute to the evolution of the SCTs. One is insertions of TPEs into the LO $T$-matrix, which, after being properly renormalized, generates an $\mathcal{O}(Q^{2})$ or $\mathcal{O}(Q^{3})$ correction to the LO [see Eq.~\eqref{eqn:LSE23}]. The structure of divergence in these diagrams has been used as the primary tool to gauge the SCTs~\cite{Long:2007vp, Valderrama:2009ei, Valderrama:2011mv, Long:2011qx}.

The other mechanism driving the SCTs is the LO amplitude itself. With the necessary counterterms, if any, the cutoff dependence of the LO amplitude vanishes at $\Lambda \to \infty$; there normally is residual cutoff dependence at finite $\Lambda$s.
When $\Lambda$ is rescaled to a smaller value, $\Lambda'$, the integrated-out momentum modes in the interval of $(\Lambda', \Lambda)$ will, for the most part, contribute to the evolution of the leading counterterm, but a perfect RG invariance would request the SCTs to evolve as well, so that there would not even be a small cutoff dependence. References~\cite{Birse:2005um, Birse:2009my} studied this mechanism using Wilson's RG equation, but their assumption about the fixed-point solutions to the RG equation seems to be at odds with the limit-cycle-like behavior of the LO counterterms of the attractive channels~\cite{Nogga:2005hy}.

A simpler way to gauge the LO-amplitude-induced SCTs is to consider how the LO residual cutoff dependence scales against $\Lambda$. According to Ref.~\cite{PavonValderrama:2007nu}, the LO residual cutoff dependence in the attractive triplet channels is $\mathcal{O}(\Lambda ^{-5/2})$, which means our ignorance of the LO-amplitude-induced SCTs is smaller than $\mathcal{O}(Q^2/\mhi^2)$---the corrections brought by TPE0 and its primordial counterterms. As a consequence, there is no need to have a nonvanishing $\mathcal{O}(Q)$ counterterm that is not accompanied by any long-range force.

Yet another evidence of vanishing $\mathcal{O}(Q)$, though \textit{a posteriori}, is that the LO error of the $\csd$ mixing angle scales as $k^2$ instead of $k$, $|\epsilon^{\text{EFT}}_1 - \epsilon^{\text{PWA}}_1| \propto k^2$, as seen in Fig. 6 of Ref.~\cite{Beane:2001bc}.

In conclusion, the counting of the SCTs in the triplet channels will be decided by the larger one of (i) their primordial size $\mathcal{O}(Q^2)$, and (ii) what the divergence of one insertion of TPE requires.

With $\mathcal{O}(Q)$ vanishing, the on-shell $\mathcal{O}(Q^{2})$ and $\mathcal{O} (Q^{3})$ $T$-matrices, $T^{(2)}$ and $T^{(3)}$, are calculated, respectively, by one insertion of $\mathcal{O}(Q^{2})$ and $\mathcal{O}(Q^{3})$ potentials into $T^{(0)}$,
\begin{equation}
\begin{split}
T^{(2,\, 3)}(k, k) & = V^{(2,\, 3)}(k, k) + \frac{4}{\pi} m_N \int^\Lambda d\kappa\, \kappa^2\, V^{(2,\, 3)}(k, \kappa) \frac{T^{(0)}(\kappa, k)}{k^2 - \kappa^2 + i\epsilon} \\
&{} \quad + \frac{4}{\pi^2}m_N^2 \int^\Lambda\int^\Lambda d\kappa\, d\kappa'\, \kappa^2\, {\kappa'}^2 \frac{T^{(0)}(k, \kappa)}{k^2 - \kappa^2 + i\epsilon} V^{(2,\, 3)}(\kappa, \kappa') \frac{T^{(0)}(\kappa', k)}{k^2 - {\kappa'}^2 + i\epsilon} \, ,
\end{split}\label{eqn:LSE23}
\end{equation} 
where $T$-matrices and $V$s are understood as $2\times 2$ matrices for the coupled channels. This is of course nothing more than the first-order distorted-wave expansion.

Treating the subleading potentials as perturbations will no doubt break the exact unitarity of the $S$-matrix, as does any perturbation-theory-based calculation. But in a consistent power counting scheme, the violation of unitarity is of higher order. This makes it slightly nontrivial to extract the phase shifts and the mixing angles from the expanded EFT $T$-matrix. Although it has been covered in the literature, we list in the Appendix~\ref{sec:conversion} the useful formulas for convenience of reference.

\subsection{Uncoupled Channels: $\cp{3}{1}$ and $\cp{3}{0}$}

In $\cp{3}{1}$, OPE (with short-distance behavior $+1/r^3$) leads to an LO wave function exponentially suppressed near the origin: $\sim \exp[-(\alpha r)^{1/2}]$, with $\alpha$ as a positive mass scale~\cite{Vald:ope}. The exponential damping of the LO wave function would eliminate the singularity of TPE0 ($\sim 1/r^5$) and TPE1 ($\sim 1/r^6$), even without any counterterm. But, as we argued, we will not demote any counterterm with respect to WPC. Therefore, the $\mathcal{O}(Q^2)$ and $\mathcal{O}(Q^3)$ $\cp{3}{1}$ counterterms are
\begin{equation}
\begin{split}
\langle\, {\cp{3}{1}}|V_{S}^{(2,\, 3)}|\,{\cp{3}{1}}\rangle = C_{{\cp{3}{1}}
}^{(0,\, 1)}\,p'p\,.
\end{split}
\label{eqn:3p1ccpresI}
\end{equation} 
The splitting of $C_\cp{3}{1}$ into different orders does not mean that we will take more than one input for $C_\cp{3}{1}$; it only reflects the possibility that the ``bare'' values of the counterterms could be modified by the short-range core of TPEs.

Perturbative renormalization at subleading orders in $\cp{3}{0}$ was first studied in Ref.~\cite{Valderrama:2011mv} and in a parallel work of ours~\cite{Long:2011qx}. Although we have reached the same conclusion about the uncoupled channels as Ref.~\cite{Valderrama:2011mv}, we include here, for completeness, our analysis of $\cp{3}{0}$~\cite{Long:2011qx}.

After the LO amplitude is renormalized with $C_\cp{3}{0}$, the LO wave function in $\cp{3}{0}$ can be approximated near the origin in powers of $k^2$ up to a normalization factor~\cite{frank, Beane:2001bc, PavonValderrama:2007nu},
\begin{equation}
\psi_{k}^{(0)}(r) \sim \left(\frac{\lambda}{r}\right) ^{\frac{1}{4}}\left[u_{0} + k^2r^2 \sqrt{\frac{r}{\lambda}}u_{1} +
\mathcal{O}(k^{4}) \right] \, ,
\label{eqn:psi0}
\end{equation}
where $\lambda =\frac{3g_{A}^{2}m_{N}}{32\pi f_{\pi }^{2}}$, $u_{0}$ and $u_{1}$ are oscillatory functions in terms of $r/\lambda$ and $\phi$ with amplitudes $\sim 1$, and $\phi$ is the phase between the two independent solutions and is related to $C_\cp{3}{0}$. Combined with the short-range behavior of TPE0, $V_{2\pi}^{(0)} \sim 1/r^5$, and TPE1, $V_{2\pi}^{(1)} \sim 1/r^6$, the superficial divergence of one insertion of TPE is estimated on a dimensional ground~\cite{Valderrama:2011mv, Long:2011qx}:
\begin{align}
T^{(0)}_{2\pi,\,\cp{3}{0}} &= \langle \psi ^{(0)}|V_{2\pi}^{(0)}|\psi ^{(0)}\rangle_\cp{3}{0} \sim
\int_{\sim 1/\Lambda }drr^{2}|\psi ^{(0)}(r)|^{2}\frac{1}{r^{5}}  \notag \\
& \sim \alpha _{0}(\Lambda )\Lambda ^{5/2}+\beta
_{0}(\Lambda )k^{2}+\mathcal{O}(k^{4}\Lambda ^{-5/2}) ,
\label{eqn:supT2} \\
T^{(1)}_{2\pi,\,\cp{3}{0}} &= \langle \psi ^{(0)}|V_{2\pi}^{(1)}|\psi ^{(0)}\rangle_\cp{3}{0} \sim
\int_{\sim 1/\Lambda }drr^{2}|\psi ^{(0)}(r)|^{2}\frac{1}{r^{6}}  \notag \\
& \sim \alpha _{1}(\Lambda )\Lambda ^{7/2}+\beta
_{1}(\Lambda )\Lambda k^{2}+\mathcal{O}(k^{4}\Lambda ^{-3/2}) \, ,
\label{eqn:supT3}
\end{align}
where $\alpha_{0,1}(\Lambda)$ and $\beta_{0,1}(\Lambda )$ are oscillatory functions diverging slower than $\Lambda$.

The presence of two divergent terms suggests that (i) the running of the LO counterterm $C_\cp{3}{0}(\Lambda)$ needs to be corrected at higher orders, and (ii) one SCT, $D_\cp{3}{0}p' p({p'}^2 + p^2)$, needs to be enlisted.
The fact that $D_\cp{3}{0}$ arises at the same order as TPE0 leads us to arrange counterterms at $\mathcal{O}(Q^2)$ and $\mathcal{O}(Q^3)$ as follows:
\begin{equation}
\begin{split}
\langle\, {\cp{3}{0}}|V_{S}^{(2,\, 3)}|\,{\cp{3}{0}}\rangle = C_{{\cp{3}{0}}
}^{(2,\, 3)}\,p^{\prime }p+D_{{\cp{3}{0}}}^{(0,\, 1)}\,p^{\prime }p({p^{\prime}}
^{2}+p^{2})\,.
\end{split}
\label{eqn:3p0ccpresII}
\end{equation}
We see that the enhancement of $D_\cp{3}{0}$ is the same as that of $C_\cp{3}{0}$: $\mathcal{O}(\mhi^2/\mlo^2)$.

The lesson of power counting learned here coincides with the conclusion of Ref.~\cite{Long:2007vp}. Although NDA fails to prescribe a counterterm at LO, we could use $\mnda$ to determine how the SCTs scale when subleading long-range potentials are taken into account, which states that the enhancement of each SCT is the same as the LO counterterm.

\subsection{Coupled channels: $\csd$ and $\cpf$\label{sec:cc}}

In the coupled channels, the LO wave functions are dominated at short distances by the attractive subchannel. Because there are three independent on-shell $T$-matrix elements, calculating the superficial divergence of TPEs using Eqs.~\eqref{eqn:supT2} and \eqref{eqn:supT3} gives rise to six divergent terms, with two for each $T$-matrix element,
as shown in detail in Ref.~\cite{Valderrama:2011mv}. Reference~\cite{Valderrama:2011mv} proposes to use six counterterms to cancel these divergent pieces on a one-to-one basis. Although it guarantees RG invariance, lost is the regularity of power counting enjoyed by $\mnda$ in the uncoupled channels.

On the other hand, the presence of six divergent terms does not necessarily mean that one must have six counterterms to achieve RG invariance. $\mnda$ suggests that when the LO long-range potential gets $\mathcal{O}(Q^2)$ correction going from OPE to TPE0, so do the SCTs; therefore, the $\mathcal{O}(Q^2)$ SCTs in the coupled channels should be the $D$ and $E$ terms in Eqs.~\eqref{eqn:3s1gnrVS} and \eqref{eqn:3p2gnrVS}, which have two more derivatives than the LO counterterm. We propose the following power counting based on $\mnda$ for the SCTs of the coupled channels:
\begin{itemize}
\item[(i)] In $\csd$, we do not change WPC,
\begin{equation}
\begin{split}
\langle\, \csd | V_S^{(2,\, 3)} |\, \csd \rangle &=
\begin{pmatrix}
C_\cs{3}{1}^{(2,\, 3)} + D^{(0,\, 1)}_\cs{3}{1}({p'}^2 + p^2) & E^{(0,\, 1)}_\text{SD}\, p^2 \\
E^{(0,\, 1)}_\text{SD}\, {p'}^2 & 0
\end{pmatrix} \, . \\
\label{eqn:sub3s1}
\end{split}
\end{equation}
\item[(ii)] In $\cpf$, an enhancement of $\mathcal{O}(\mhi^2/\mlo^2)$ leads to
\begin{equation}
\begin{split}
\langle\, \cpf | V_S^{(2,\, 3)} |\, \cpf \rangle &= p' p
\begin{pmatrix}
C_\cp{3}{2}^{(2,\, 3)} + D^{(0,\, 1)}_\cp{3}{2}({p'}^2 + p^2) & E^{(0,\, 1)}_\text{PF}\, p^2 \\
E^{(0,\, 1)}_\text{PF}\, {p'}^2 & 0
\end{pmatrix} \, .\\
\label{eqn:sub3p2}
\end{split}
\end{equation}
\end{itemize}
An analytical proof of renormalizability with the above counterterms is difficult because the closed form of the LO $T$-matrix is not available, so we will resort to numerical experiments in Sec.~\ref{sec:numerics} to test RG invariance, or the lack thereof.

At a given energy, there are three scattering parameters---two phase shifts and one mixing angle---to be extracted from the $2\times2$ $T$-matrix \eqref{eqn:stapp}. With $T^{(2)}$ determined by three inputs from partial-wave analysis (PWA) and Eq.~\eqref{eqn:S2abc}, Eq.~\eqref{eqn:LSE23} provides a group of linear equations to solve for $C^{(2)}$, $D^{(0)}$, and $E^{(0)}$. Though not obvious, it is straightforward to show that the three linear equations built from the three PWA inputs at the same energy are linearly dependent. Therefore, the needed three PWA inputs must be incorporated from at least two different energies. Not surprisingly, the same also applies to $T^{(3)}$.

\subsection{Numerics\label{sec:numerics}}

We will use TPEs without the explicit delta-isobar to demonstrate renormalization. There are a few versions of TPEs~\cite{Ordonez:1993-1995, Kaiser:1997mw, Kaiser:1998wa, Epelbaum:1998ka-1999dj, Rentmeester:1999vw, Entem:2001cg-2002sf} in the literature with slight differences in how double counting is avoided~\cite{Friar:1999sj}. For definitiveness, we use the version in Ref.~\cite{Epelbaum:1998ka-1999dj}, i.e., delta-less TPE expressions with dimensional regularization. We adopt the following low-energy constants for the $\nu =1$ $\pi \pi NN$ seagull couplings (GeV$^{-1}$): $c_{1}=-0.81$, $c_{3}=-4.7$, and $c_{4}=3.4$~\cite{Buettiker:1999ap}. With the delta integrated out, we expect the EFT expansion to break down around $Q \sim \delta$, where $\delta  \simeq 300$ MeV is the delta-nucleon mass splitting. Such a breakdown scale is noticed in, e.g., Ref.~\cite{Birse:2007sx} through ``deconstruction.''

\begin{figure}
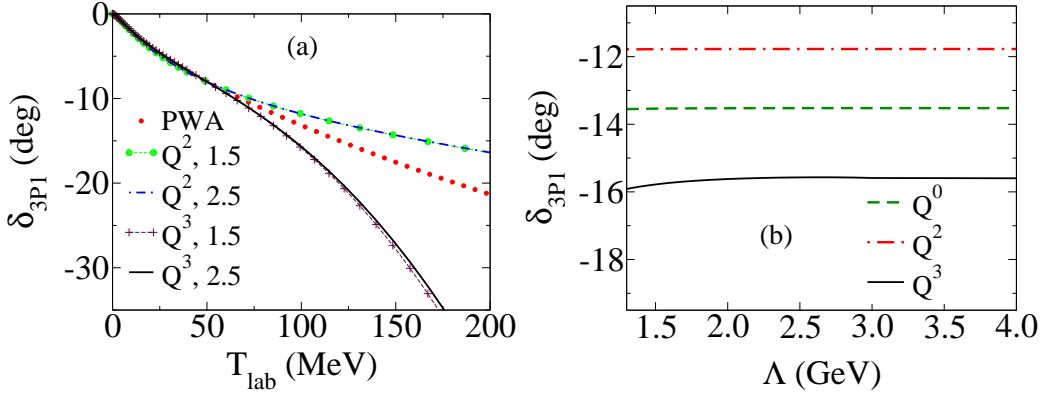

\includegraphics[scale=0.25, clip=true]{3p1-1500_mod.eps}
\includegraphics[scale=0.25, clip=true]{fxk-2-3p1_mod.eps}
\caption{(Color online) With the SCTs \eqref{eqn:3p1ccpresI}, the $\cp{3}{1}$ phase shifts (a) as a function of $T_\text{lab}$ up to $\mathcal{O}(Q^3)$, and (b) as a function of $\Lambda$ for $T_\text{lab} = 100$ MeV. In the legend of (a), the number following the symbols is the cutoff value in GeV. The red dots are from the Nijmegen PWA.} 
\label{fig:3p1}
\end{figure}

Figure~\ref{fig:3p1} shows the $\cp{3}{1}$ phase shifts as a function of laboratory kinetic energy $T_\text{lab}$ for two different cutoffs and a function of the cutoff at $T_\text{lab} = 100$ MeV. The sharp momentum cutoff is chosen as the regulator throughout the paper: $\theta(\Lambda - \kappa)$, where $\kappa$ is the magnitude of the loop momentum, as defined in Eq.~\eqref{eqn:LSE0}. The values of $C_\cp{3}{1}^{(0,\, 1)}$ are solved for by a fit of the phase shift to the Nijmegen PWA~\cite{Stoks:1993tb} for $T_\text{lab} = 50$ MeV. The cutoff independence is not surprising since the LO wave function is exponentially suppressed at short distances. The large shift from $\mathcal{O}(Q^2)$ to $\mathcal{O}(Q^3)$ indicates the influence of the uncertainties of $\pi \pi NN$ coupling constants $c_i$. Nevertheless, a decent agreement with the PWA up to $T_\text{lab} = 100$ MeV is obtained.

$\cp{3}{0}$ has been thoroughly studied in our momentum-space framework in Ref.~\cite{Long:2011qx}, which confirmed the RG invariance of $\mnda$ for the uncoupled channels; see \eqref{eqn:3p0ccpresII}. We refer the reader to Ref.~\cite{Long:2011qx} for the numerical results.

\begin{figure}
\includegraphics[scale = 0.45, clip = true]{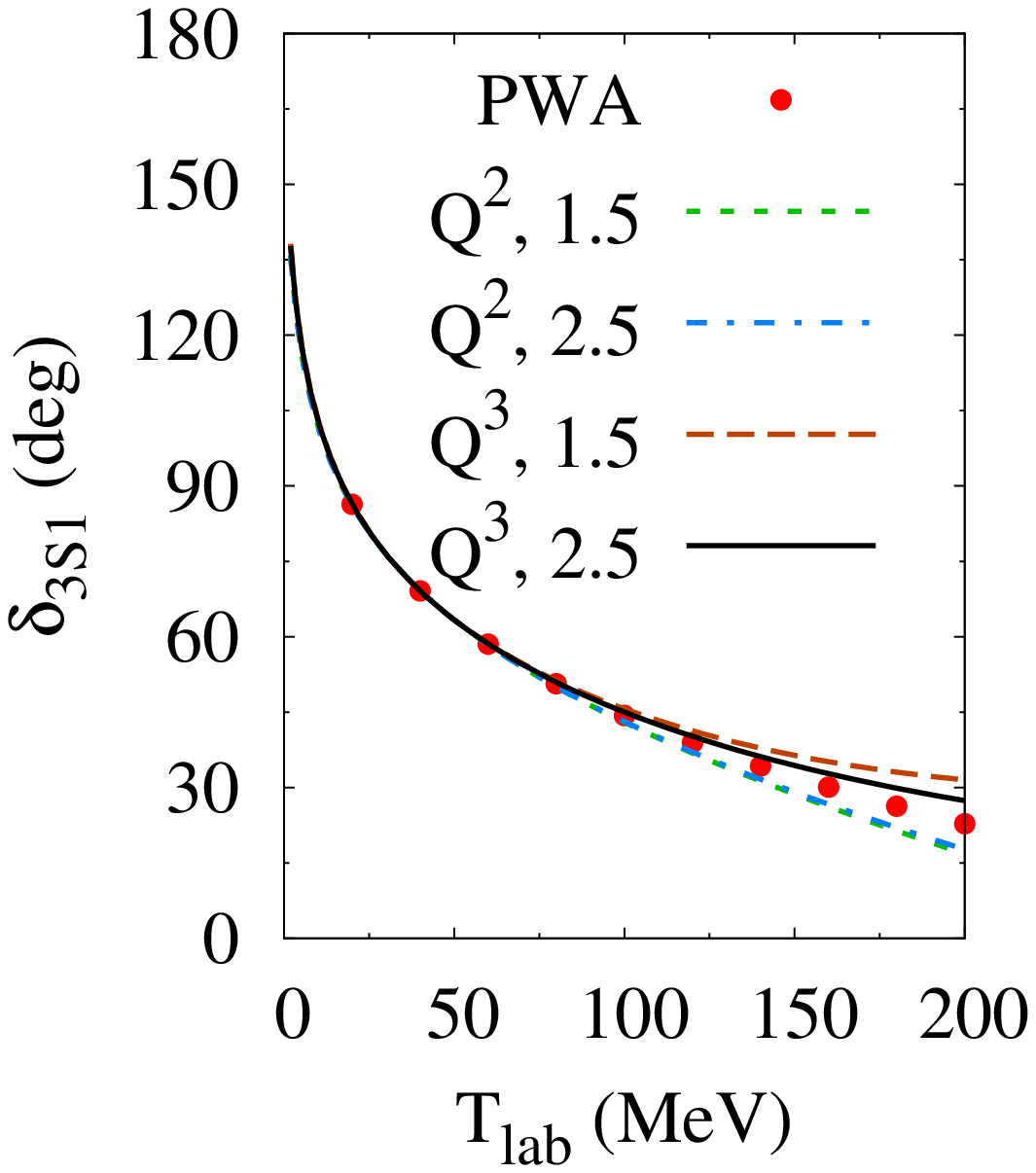}
\includegraphics[scale = 0.45, clip = true]{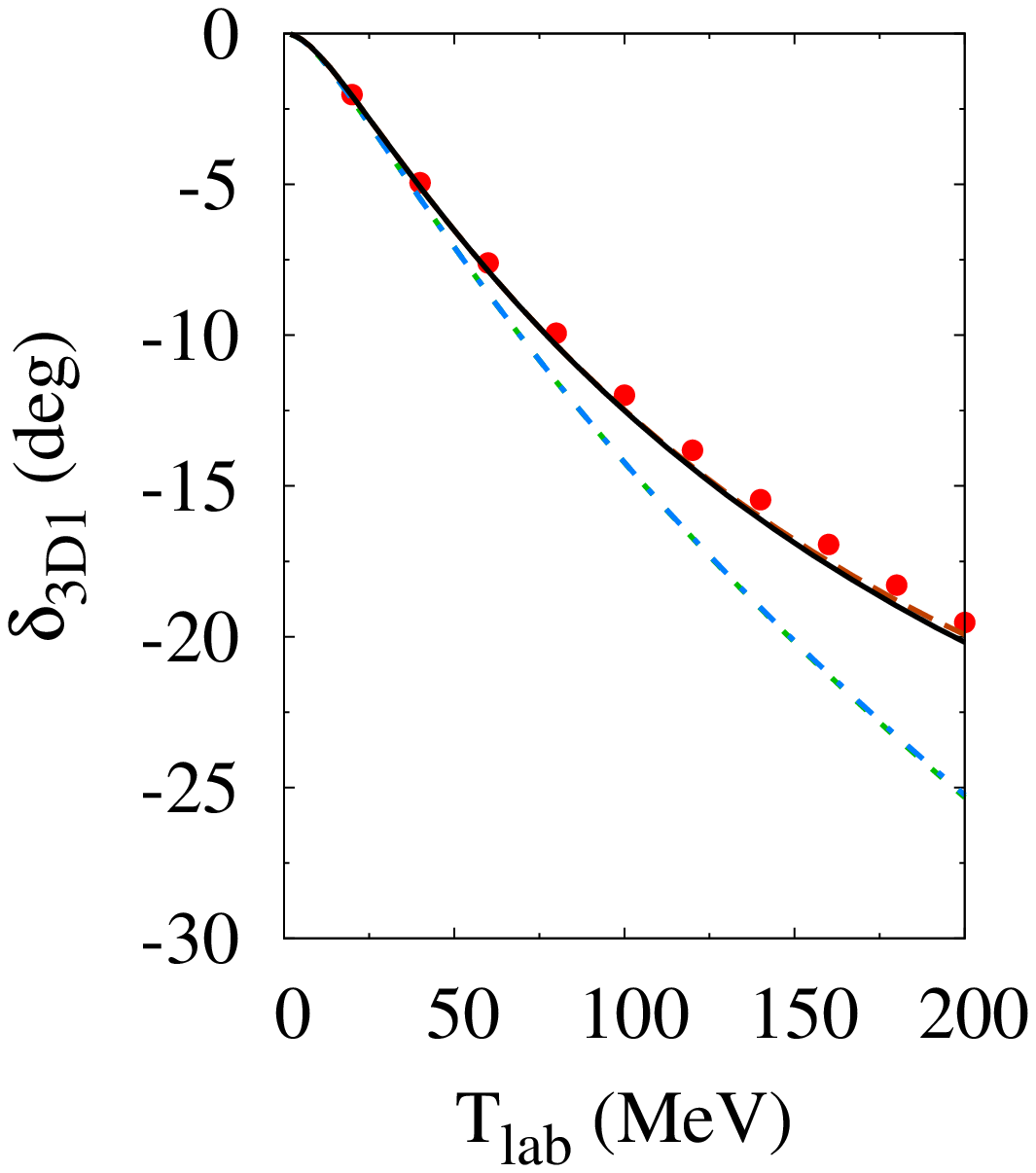}
\includegraphics[scale = 0.45, clip = true]{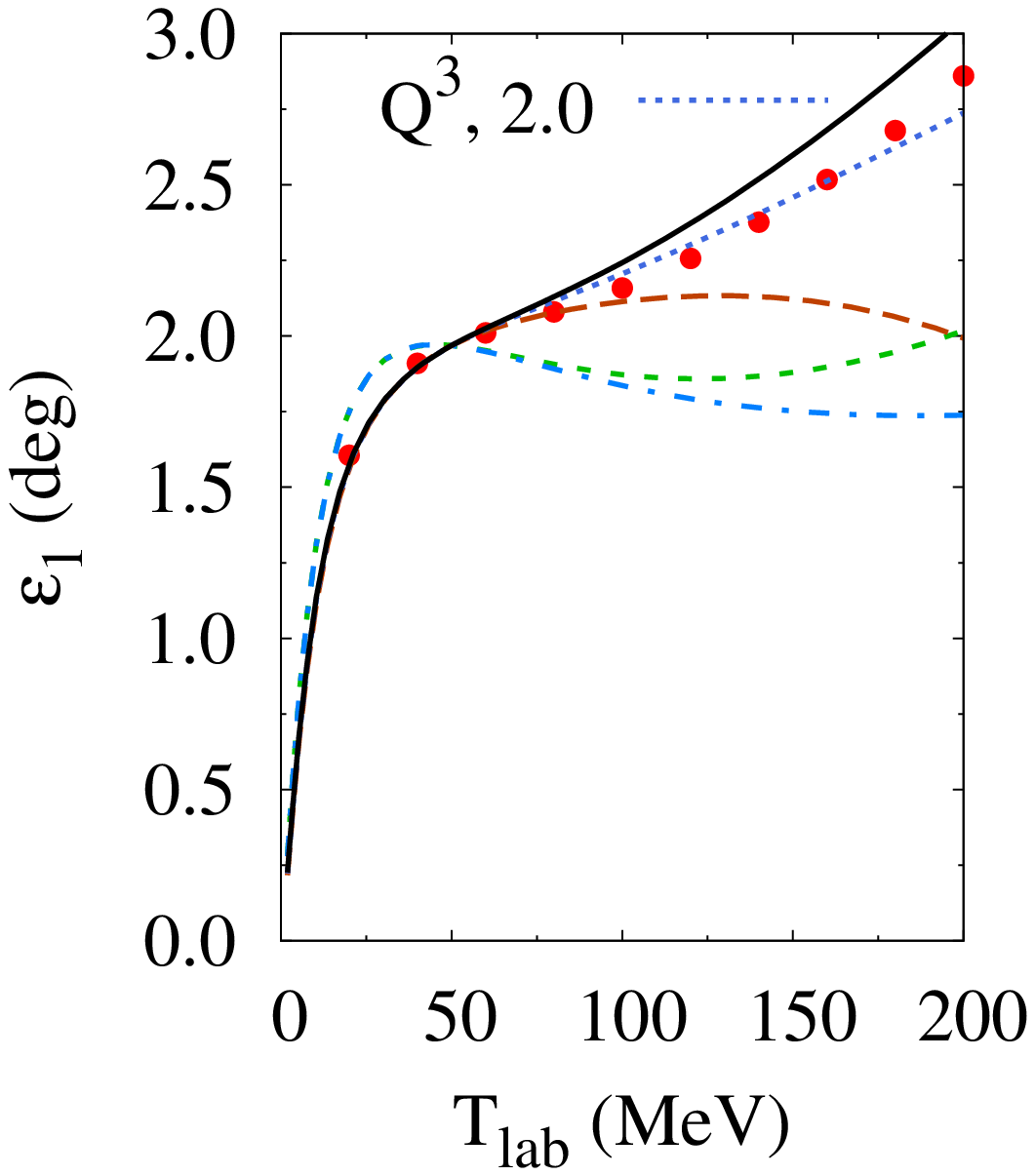}
\caption{(Color online) With the SCTs \eqref{eqn:sub3s1}, the $\cs{3}{1}$, $\cd{3}{1}$ phase shifts and the mixing angle $\epsilon_1$ as functions of $T_\text{lab}$ at $\mathcal{O}(Q^2)$ and $\mathcal{O}(Q^3)$. In the legend, the number in front of the line symbols is the cutoff value in GeV.}
\label{fig:tlab3s1}
\end{figure}

Now we move on to the coupled channels. Figure~\ref{fig:tlab3s1} shows the phase shifts of $\csd$ and the mixing angle $\epsilon_1$ as functions of $T_\text{lab}$ at $\mathcal{O}(Q^2)$ and $\mathcal{O}(Q^3)$. The values of the counterterms are determined such that the EFT curves reproduce the Nijmegen PWA for $\delta_\cs{3}{1}$ at $T_\text{lab} = 30$ and $50$ MeV and $\epsilon_1$ at $50$ MeV. Consequently, the $\cd{3}{1}$ EFT phase shifts are predictions. The first indication of the cutoff independence is the closeness of two EFT curves with $\Lambda = 1.5$ and $2.5$ GeV. The plot of $\epsilon_1$ shows a larger cutoff dependence toward higher energies, but the fact that the $\Lambda = 2.0$ GeV curve is closer to $\Lambda = 2.5$ GeV than $\Lambda = 1.5$ GeV suggests that the cutoff independence is finally achieved at larger $\Lambda$s.

\begin{figure}[tbp]
\centering
\begin{tabular}{rr}
\includegraphics[scale=0.53, clip=true]{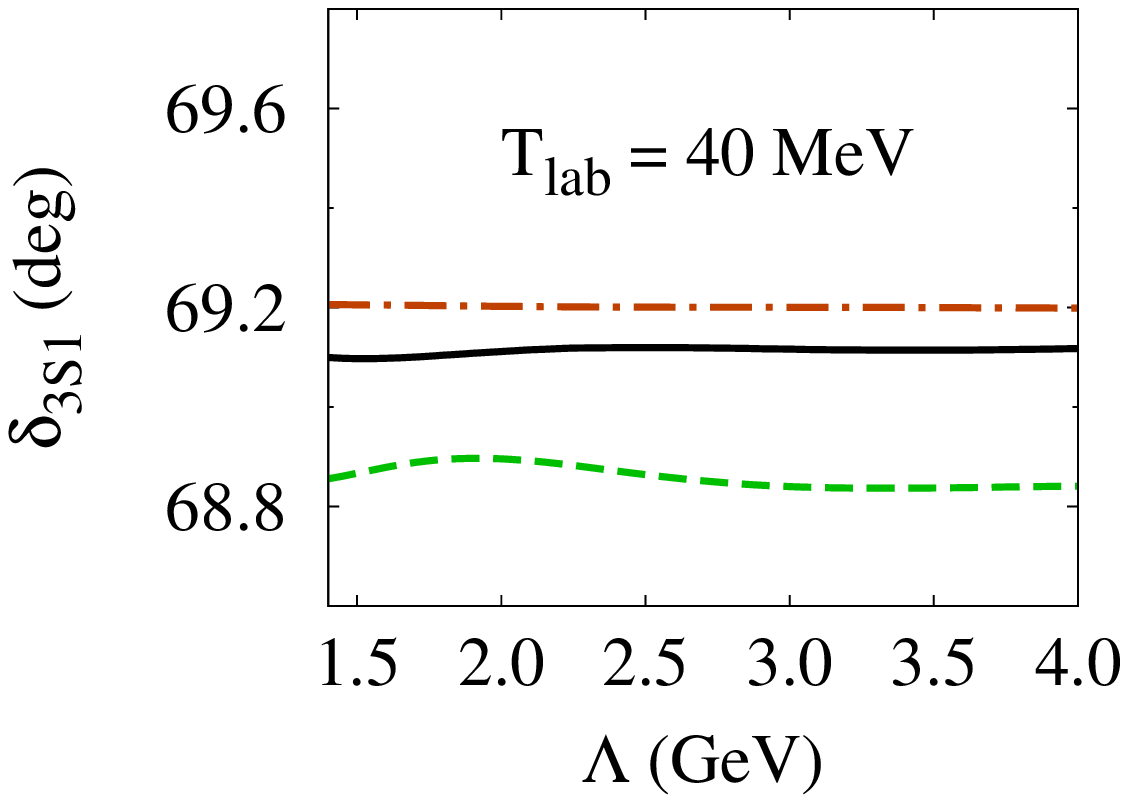} & \hspace{5mm}
\includegraphics[scale=0.53, clip=true]{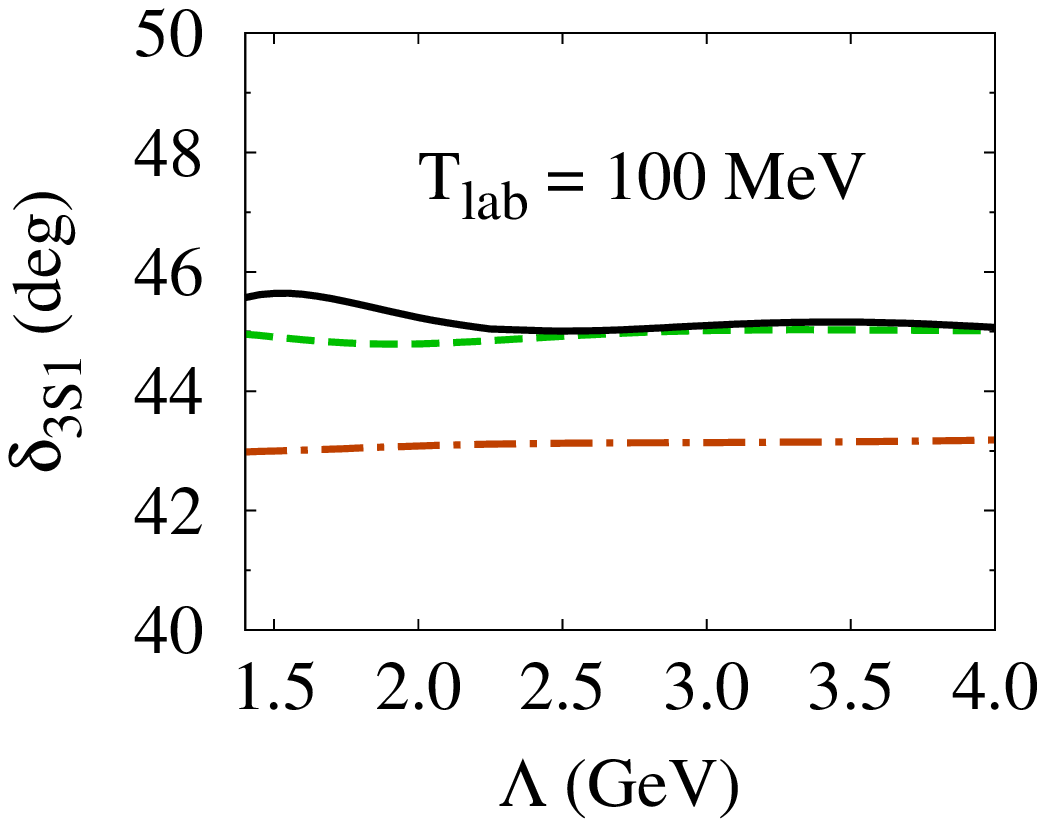}\\
\includegraphics[scale=0.53, clip=true]{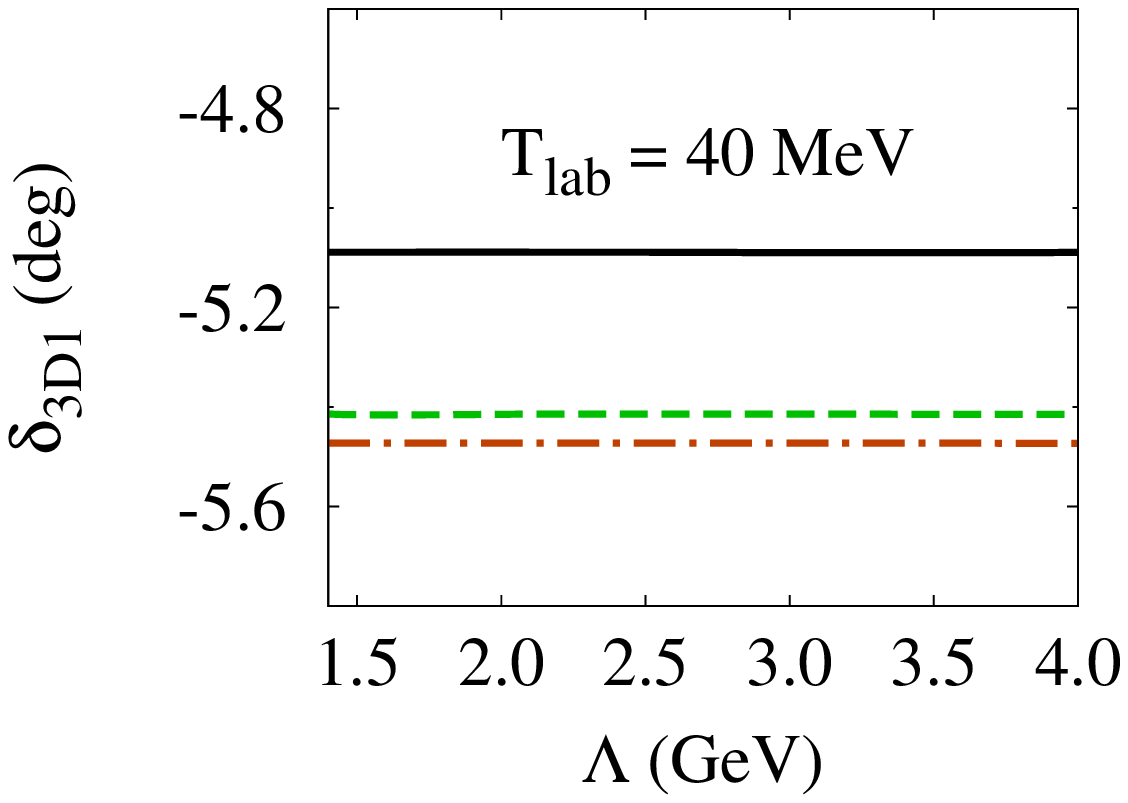} & \hspace{5mm}
\includegraphics[scale=0.53, clip=true]{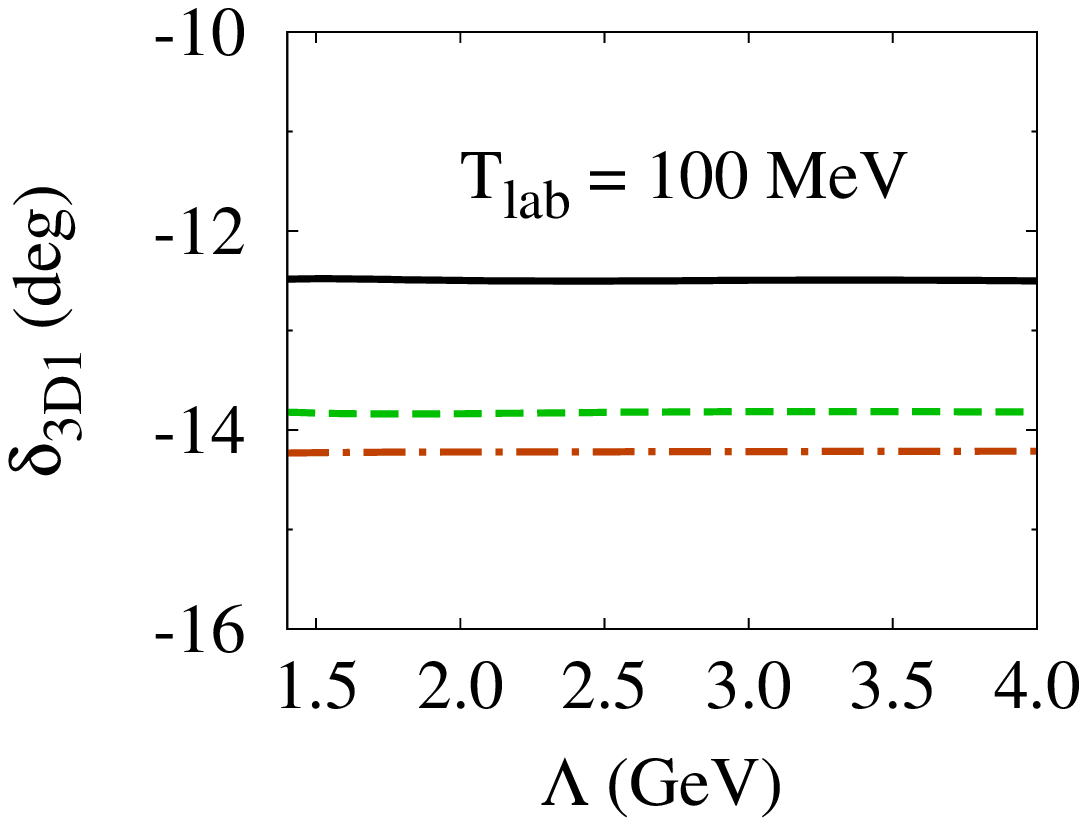}\\
\includegraphics[scale=0.53, clip=true]{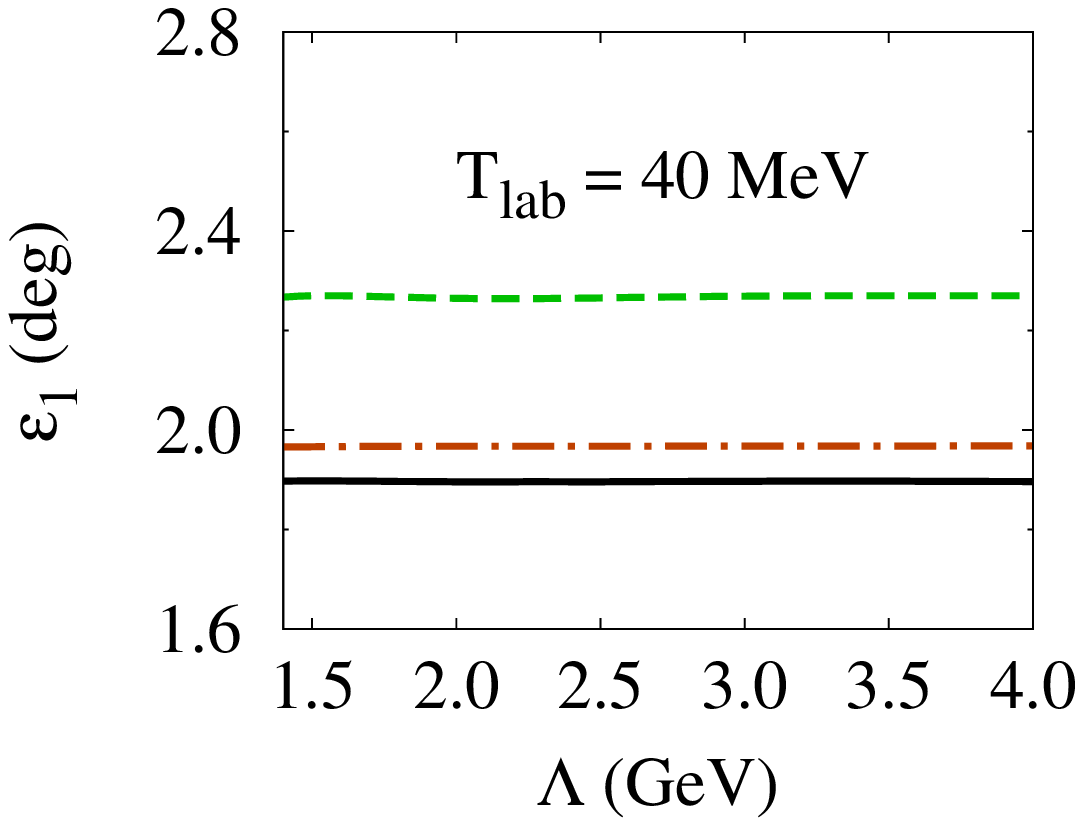} & \hspace{5mm}
\includegraphics[scale=0.53, clip=true]{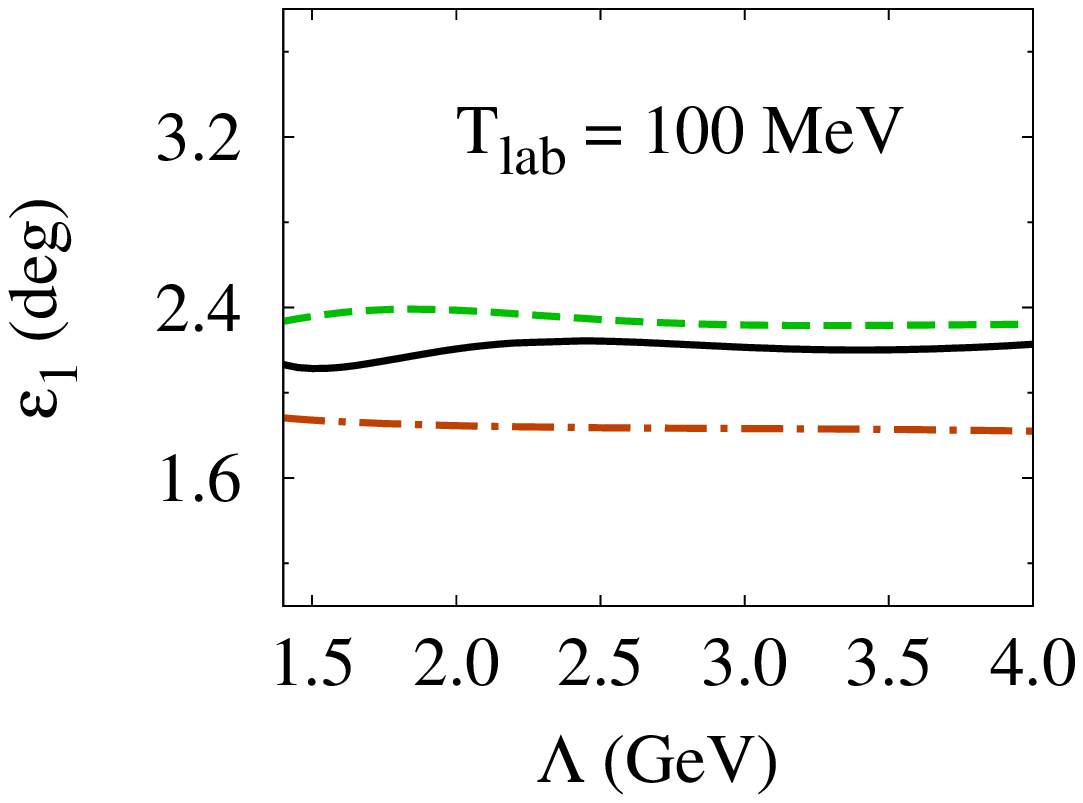}
\end{tabular}
\caption{(Color online) With the SCTs \eqref{eqn:sub3s1}, the $\cs{3}{1}$, $\cd{3}{1}$ phase shifts and the mixing angle $\epsilon_1$ at $T_\text{lab} = 40$ and $100$ MeV, as functions of the momentum cutoff. The dashed, dot-dashed and solid lines are $\mathcal{O}(1)$, $\mathcal{O}(Q^2)$ and $\mathcal{O}(Q^3)$, respectively.}
\label{fig:fxk-3s1-pres3_012_all}
\end{figure}

The cutoff independence is more clearly demonstrated in Fig.~\ref{fig:fxk-3s1-pres3_012_all}, which shows the phase shifts and the mixing angle at $T_\text{lab} = 40$ and $100$ MeV as functions of $\Lambda$. The residual cutoff dependence is still visible at lower $\Lambda$s, but it is much smaller than the size of the corresponding EFT correction. That is, the corrections are meaningful even at the lower cutoffs because they are not washed out by the cutoff uncertainties.

\begin{figure}[tbp]
\includegraphics[scale = 0.45, clip = true]{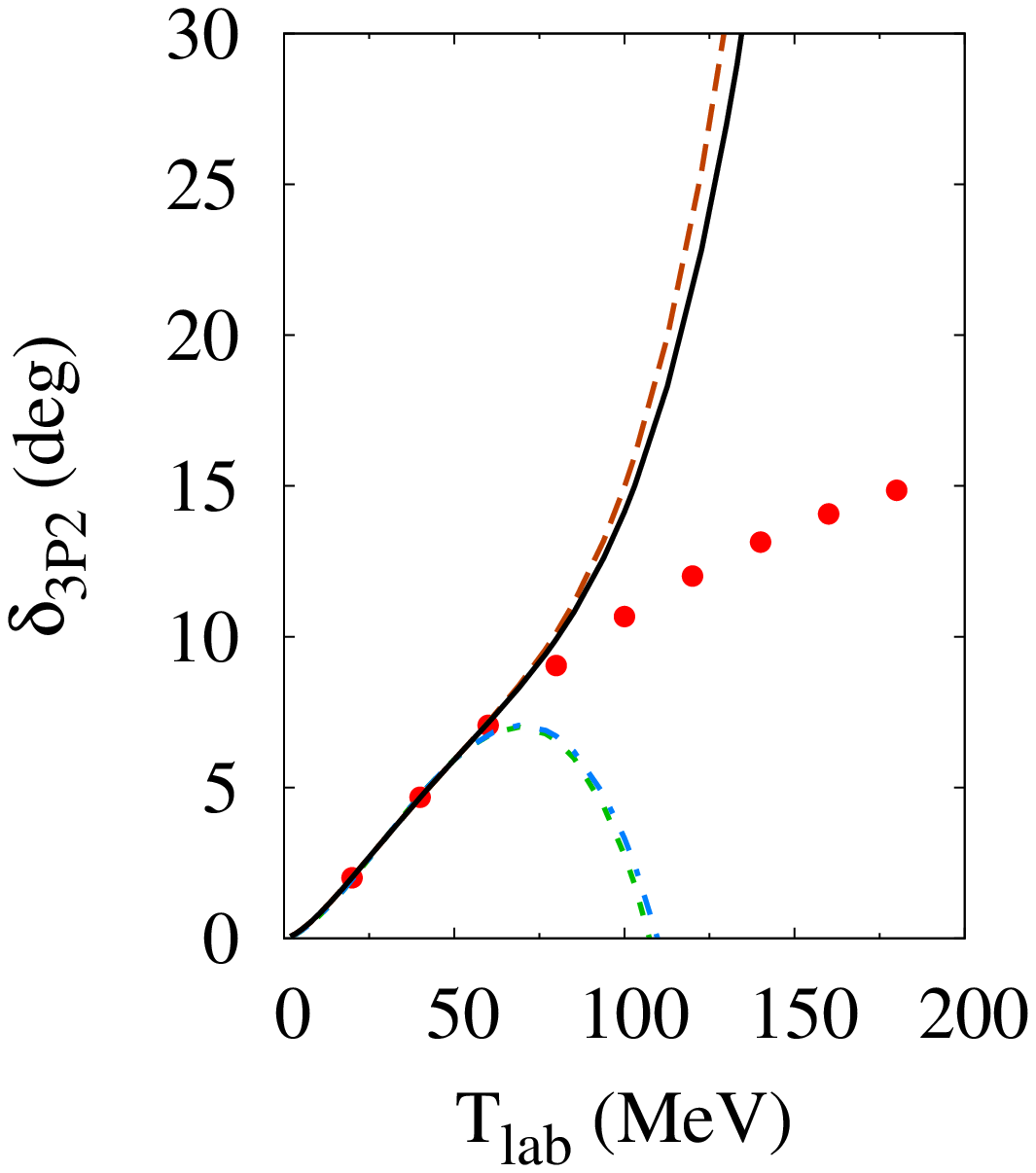}
\includegraphics[scale = 0.45, clip = true]{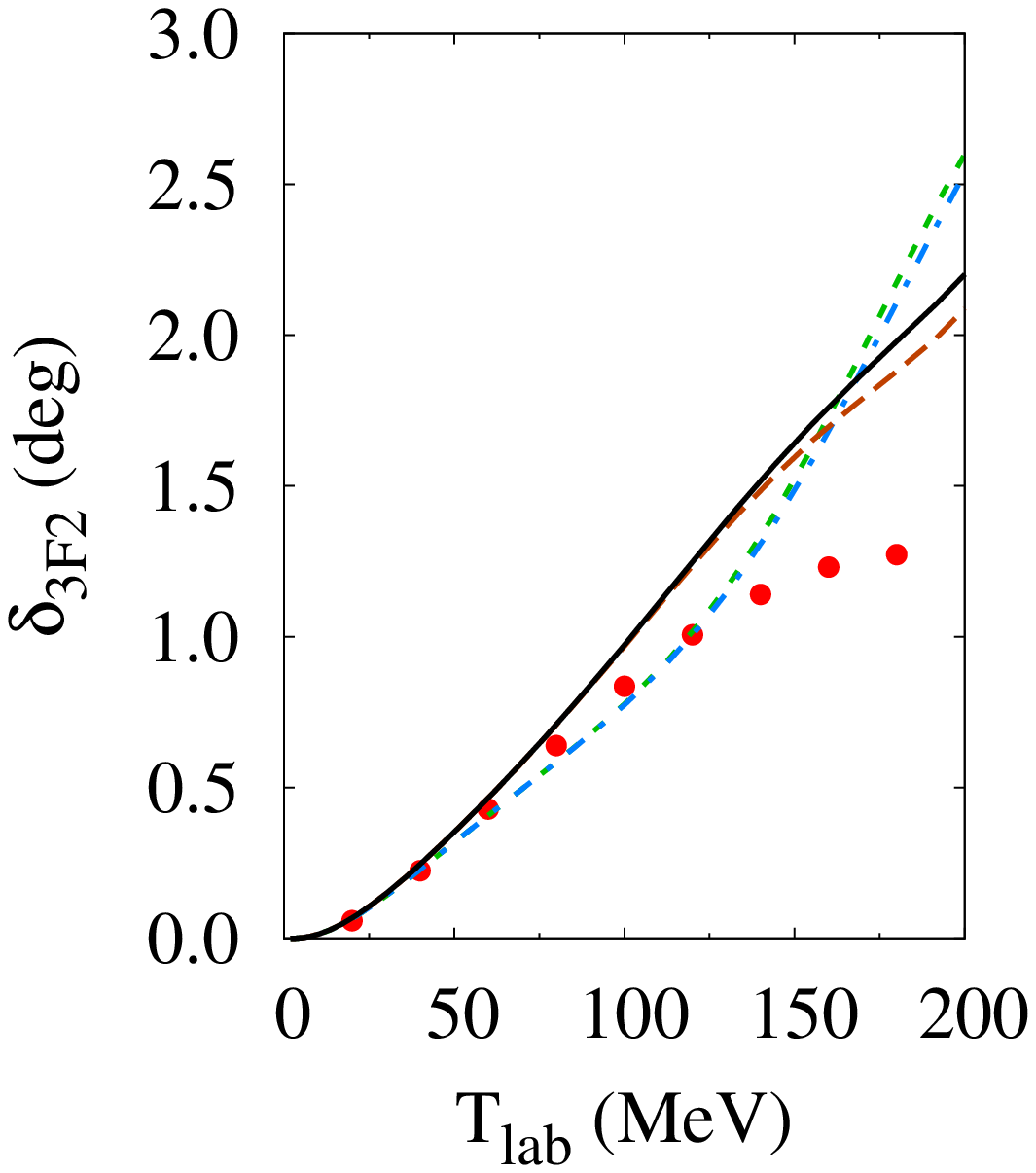}
\includegraphics[scale = 0.45, clip = true]{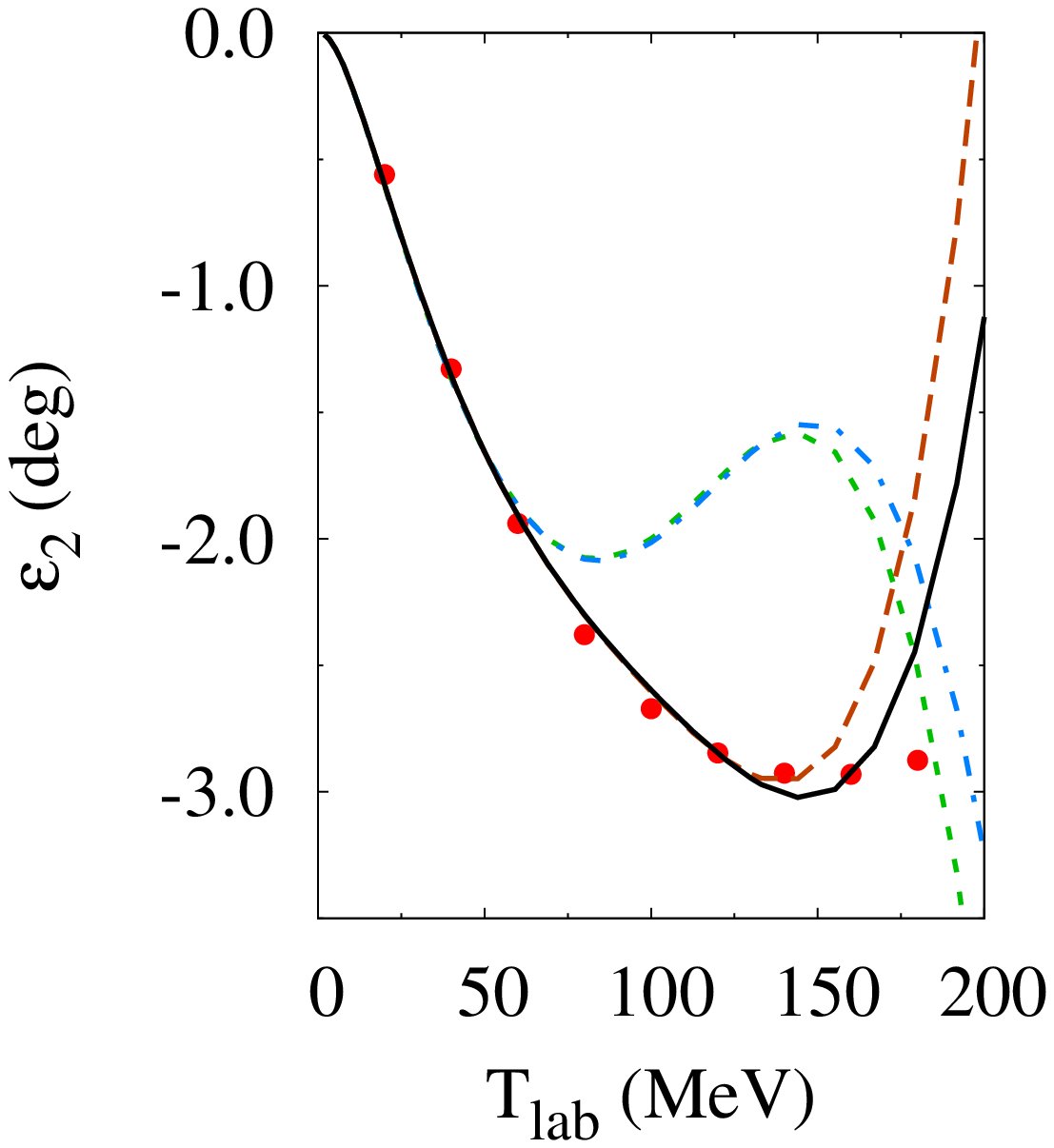}
\caption{(Color online) With the SCTs \eqref{eqn:sub3p2}, the $\cp{3}{2}$, $\cf{3}{2}$ phase shifts and the mixing angle $\epsilon_2$ as functions of $T_\text{lab}$ at $\mathcal{O}(Q^2)$ and $\mathcal{O}(Q^3)$. The symbols are explained in the legend of Fig.~\ref{fig:tlab3s1}.}
\label{fig:tlab3p2}
\end{figure}

\begin{figure}[tbp]
\centering
\begin{tabular}{rr}
\includegraphics[scale=0.53, clip=true]{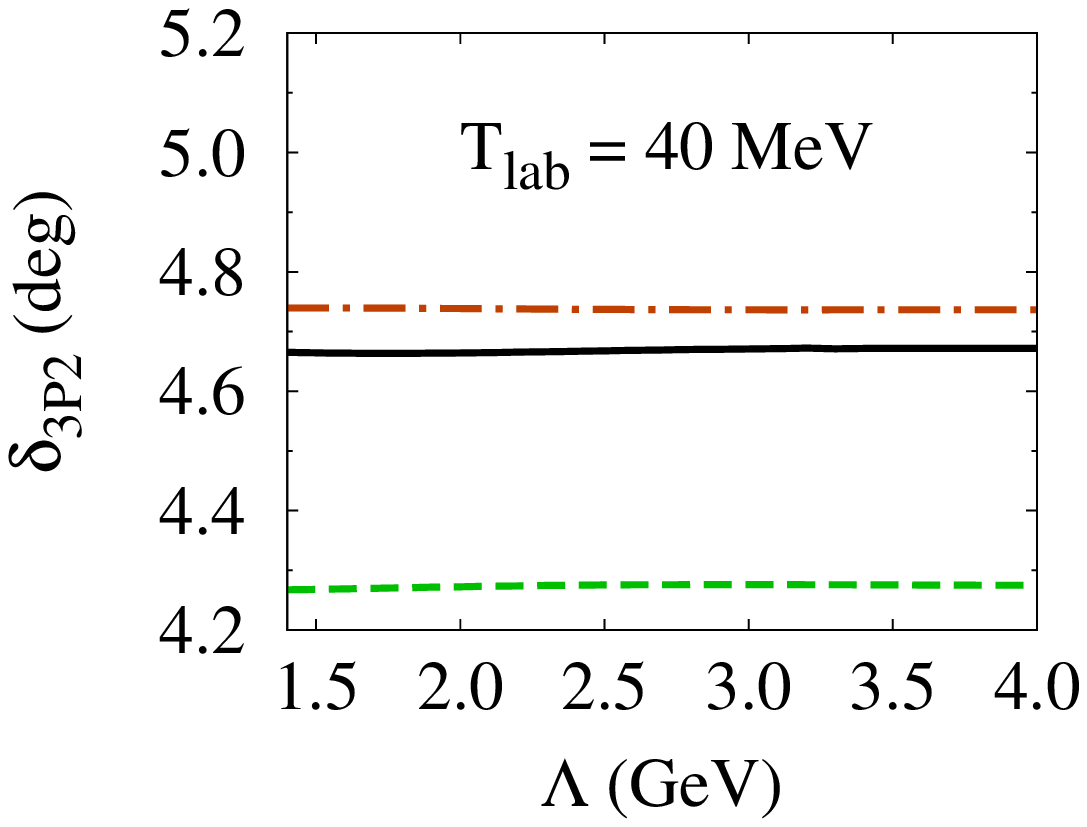} & \hspace{5mm}
\includegraphics[scale=0.53, clip=true]{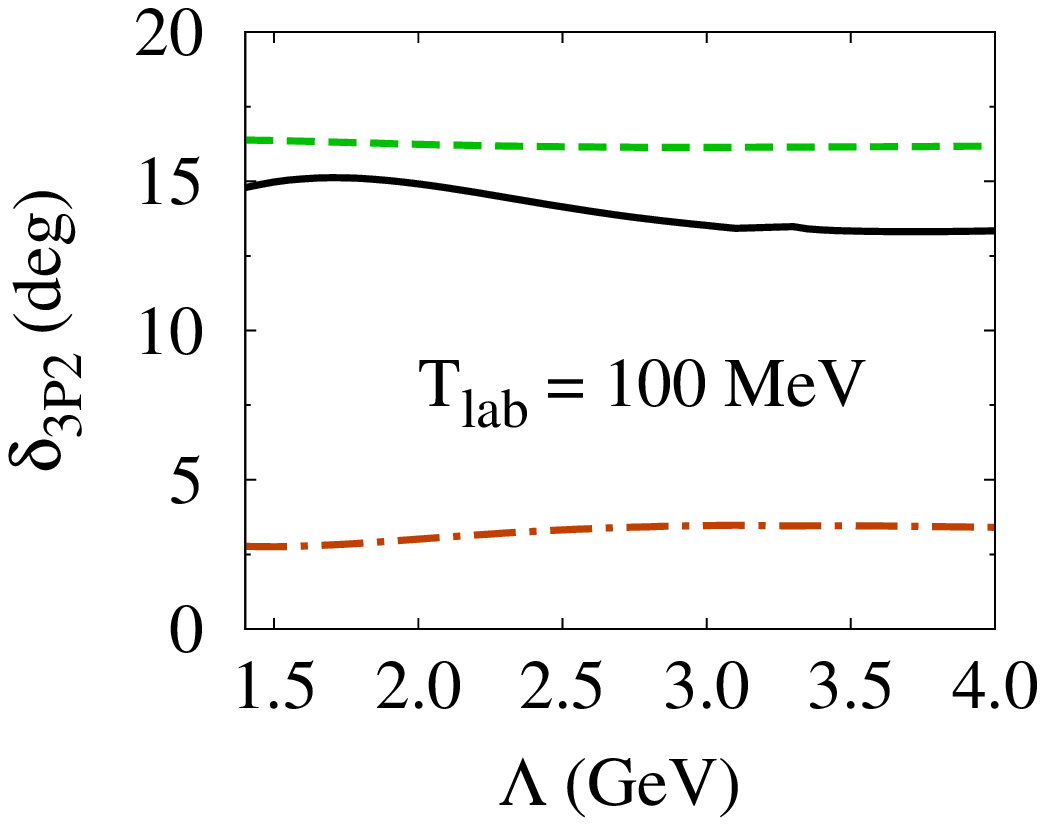}\\
\includegraphics[scale=0.53, clip=true]{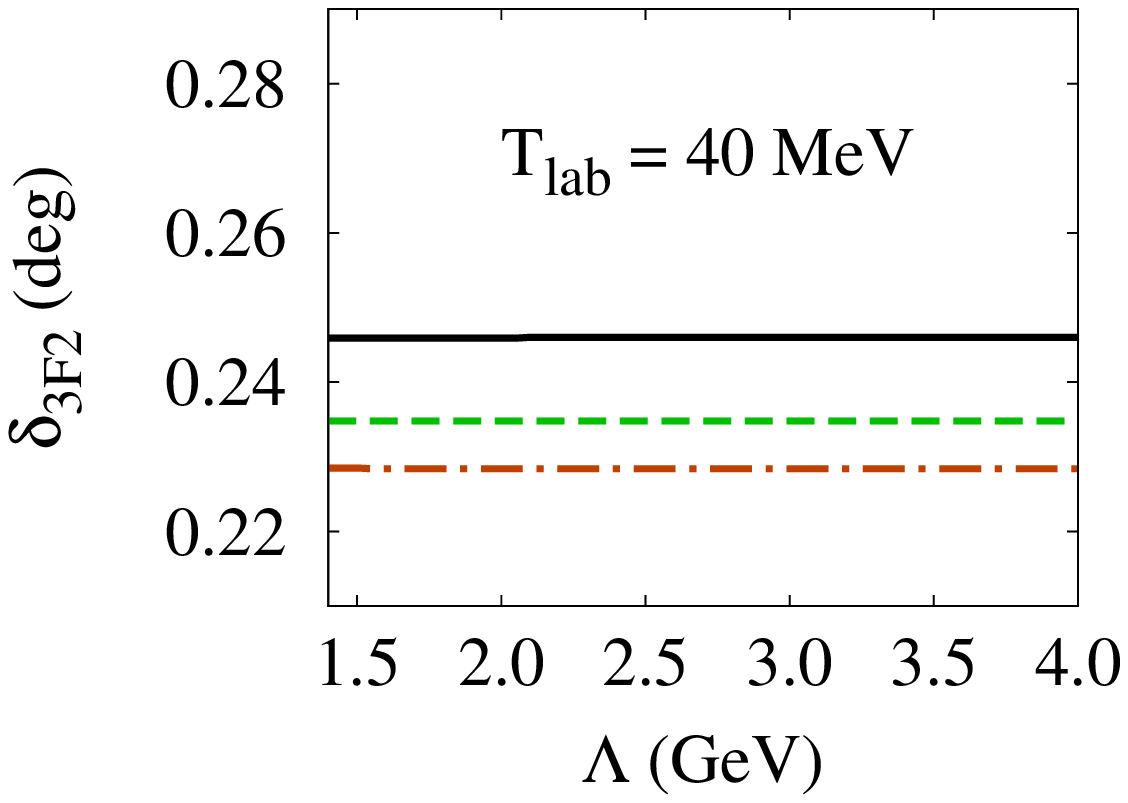} & \hspace{5mm}
\includegraphics[scale=0.53, clip=true]{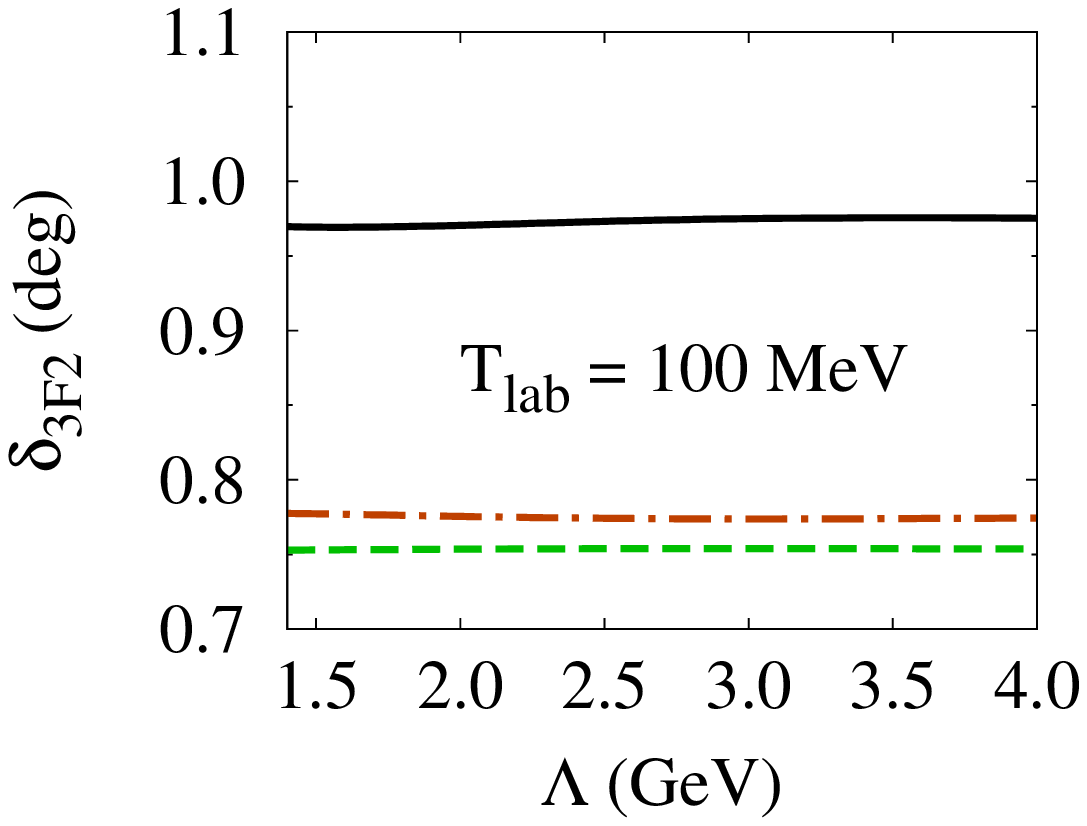}\\
\includegraphics[scale=0.53, clip=true]{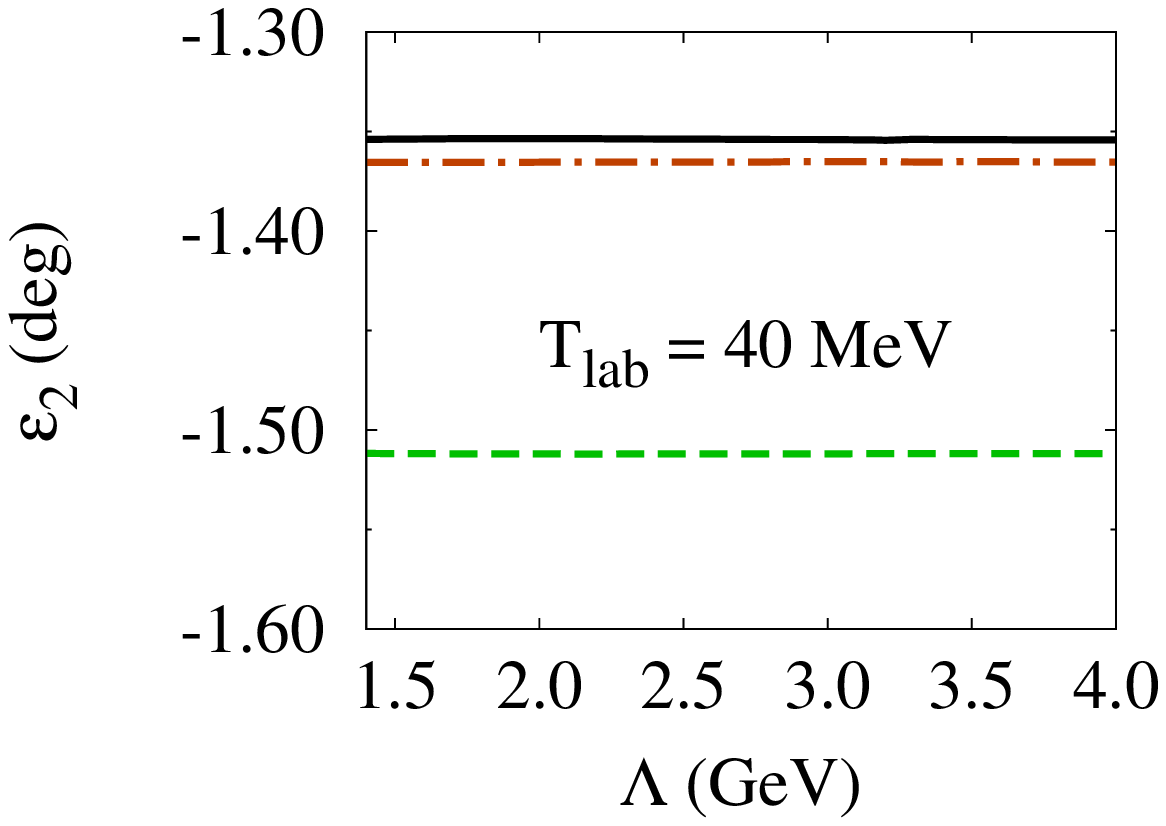} & \hspace{5mm}
\includegraphics[scale=0.53, clip=true]{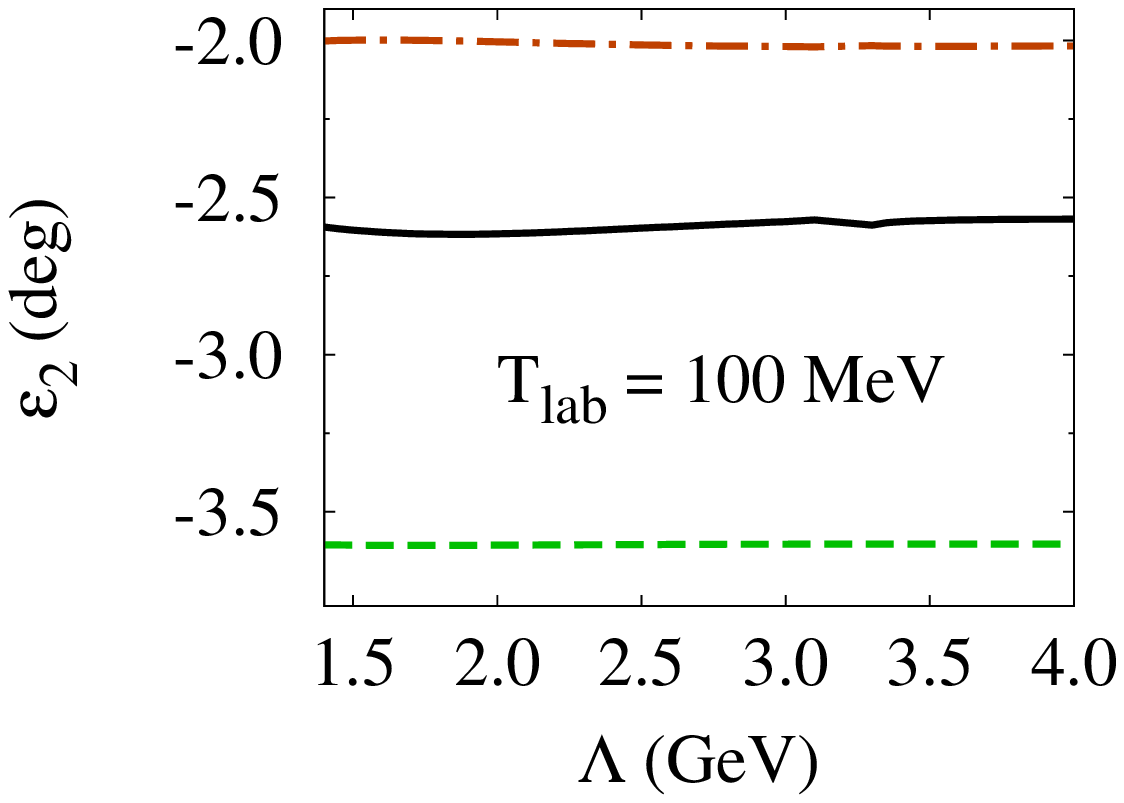}
\end{tabular}
\caption{(Color online) With SCTs \eqref{eqn:sub3p2}, the $\cp{3}{2}$, $\cf{3}{2}$ phase shifts and the mixing angle $\epsilon_2$ at $T_\text{lab} = 40$ and $100$ MeV, as functions of the momentum cutoff. The symbols are explained in the caption of Fig.~\ref{fig:fxk-3s1-pres3_012_all}.}
\label{fig:fxk-3p2-pres3_012_all}
\end{figure}

Figures~\ref{fig:tlab3p2} and \ref{fig:fxk-3p2-pres3_012_all} show the $\cpf$ phase shifts and the mixing angle $\epsilon_2$ at $\mathcal{O}(Q^2)$ and $\mathcal{O}(Q^3)$ as functions of $T_\text{lab}$ and the cutoff, respectively, with the SCTs \eqref{eqn:sub3p2}. Similar to the case of $\csd$, we fit $\delta_\cp{3}{2}$ at $T_\text{lab} = 30$ and $50$ MeV and $\epsilon_2$ at 50 MeV to the PWA values.

The main goal of this paper is to verify the RG invariance of our power counting in the UV region, so the cutoff window was chosen such that the EFT curves start to show the cutoff independence. For smaller cutoffs not shown in the plots ($1.2 \gtrsim \Lambda \gtrsim 0.6\,\text{GeV}$), the general trend is similar to $\cp{3}{0}$ (Fig.~2 of Ref.~\cite{Long:2011qx}): the $\mathcal{O}(Q^2)$ EFT curve is the first to become cutoff independent, while the LO is the latest.

A good fit to the PWA up to $100$ MeV is presented at $\mathcal{O}(Q^3)$ in $\csd$. Without special effort to improve the fits, the EFT result agrees less well with the PWA in $\cpf$. We think that this is largely owing to a disappointing LO, which departs quickly from the PWA as the energy increases. The unusually small $\cp{3}{2}$ scattering volume, $\alpha_\cp{3}{2} \simeq -0.28$ fm$^3$, compared with $\alpha$ of other $P$ waves, $|\alpha| \simeq 1.5 - 2.8$ fm$^3$~\cite{PavonValderrama:2005ku}, is suggestive of a certain amount of fine tuning, which calls for a more sophisticated fitting strategy.

\begin{figure}[tbp]
\includegraphics[scale = 0.45, clip = true]{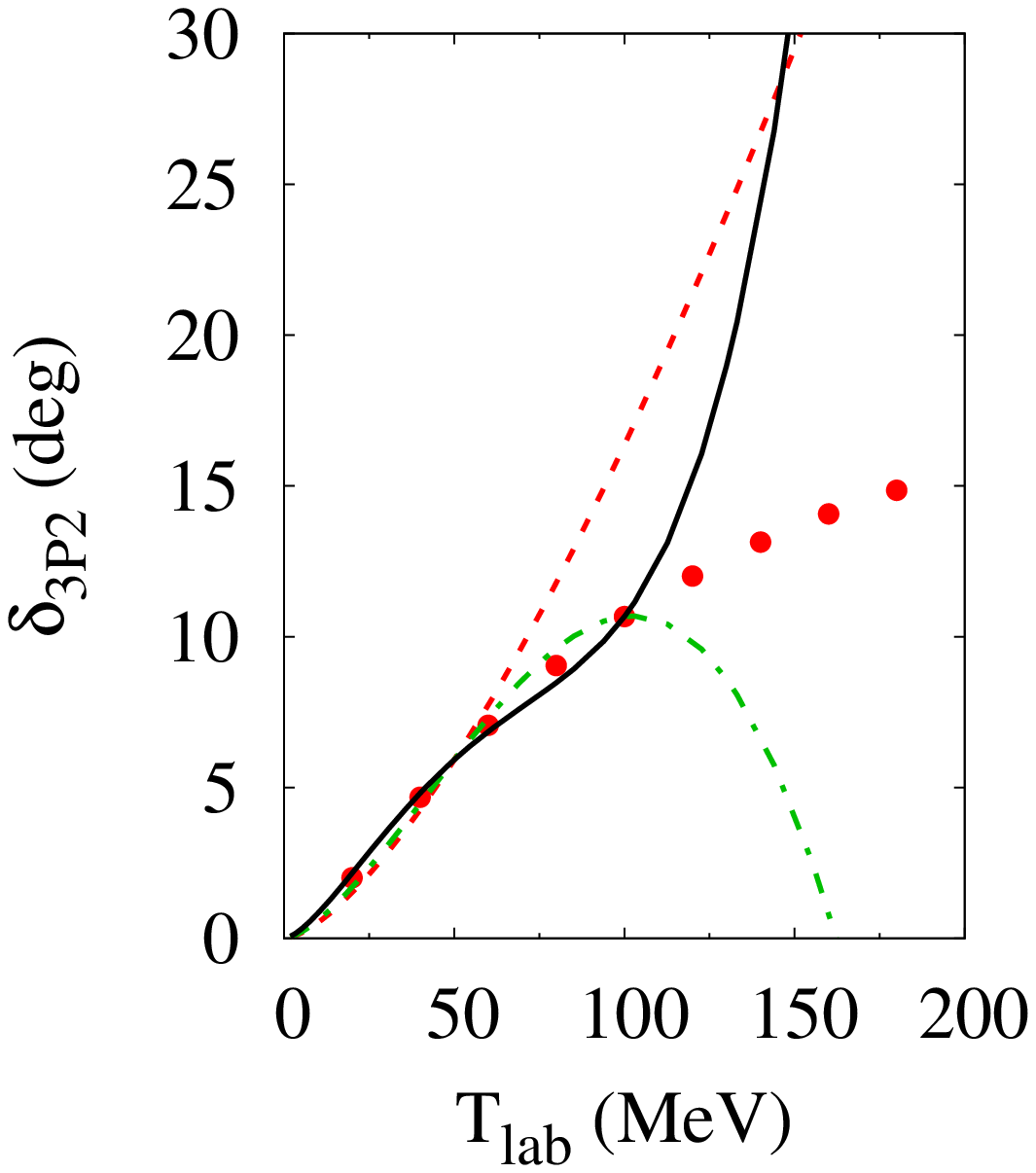}
\includegraphics[scale = 0.45, clip = true]{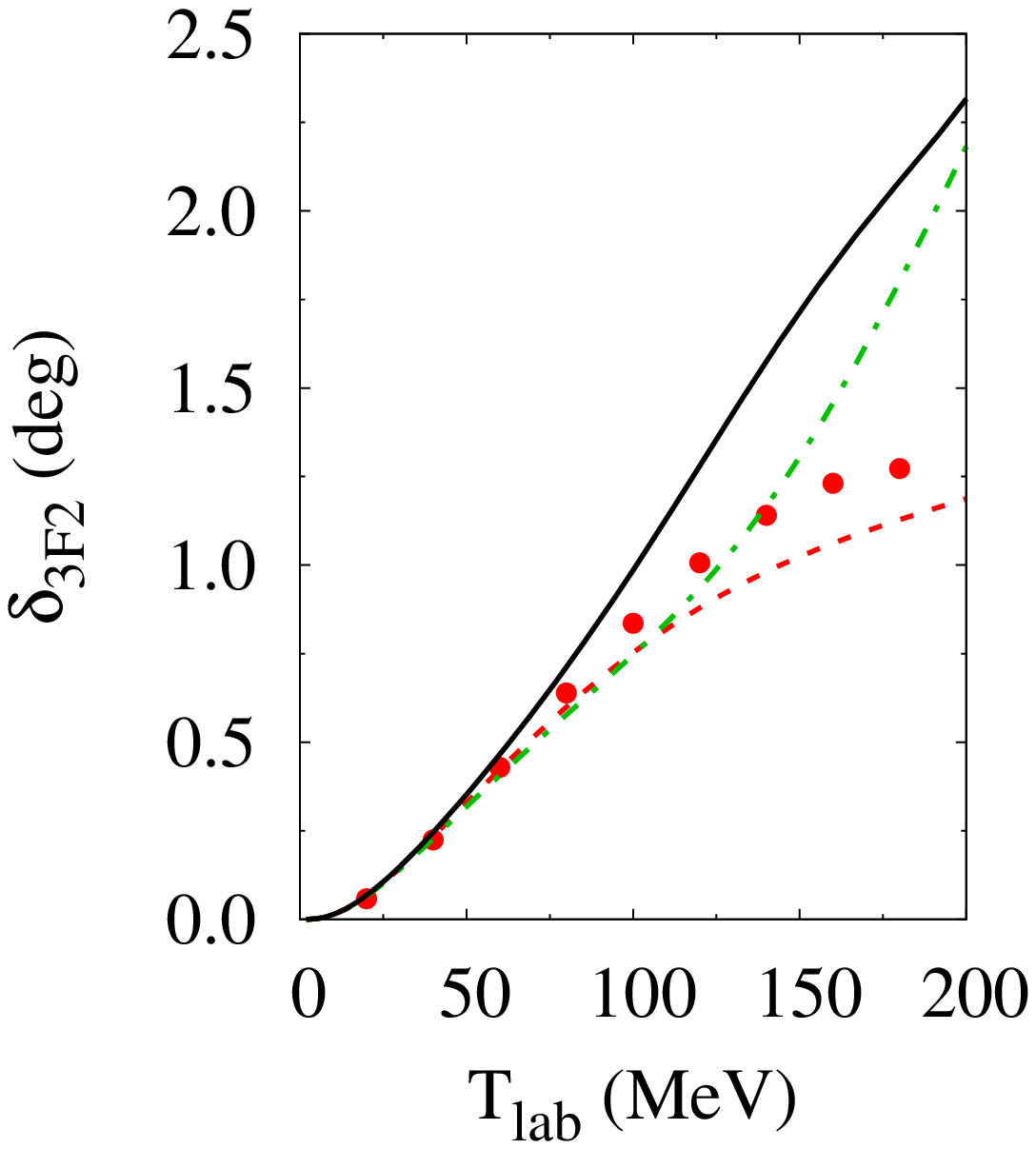}
\includegraphics[scale = 0.45, clip = true]{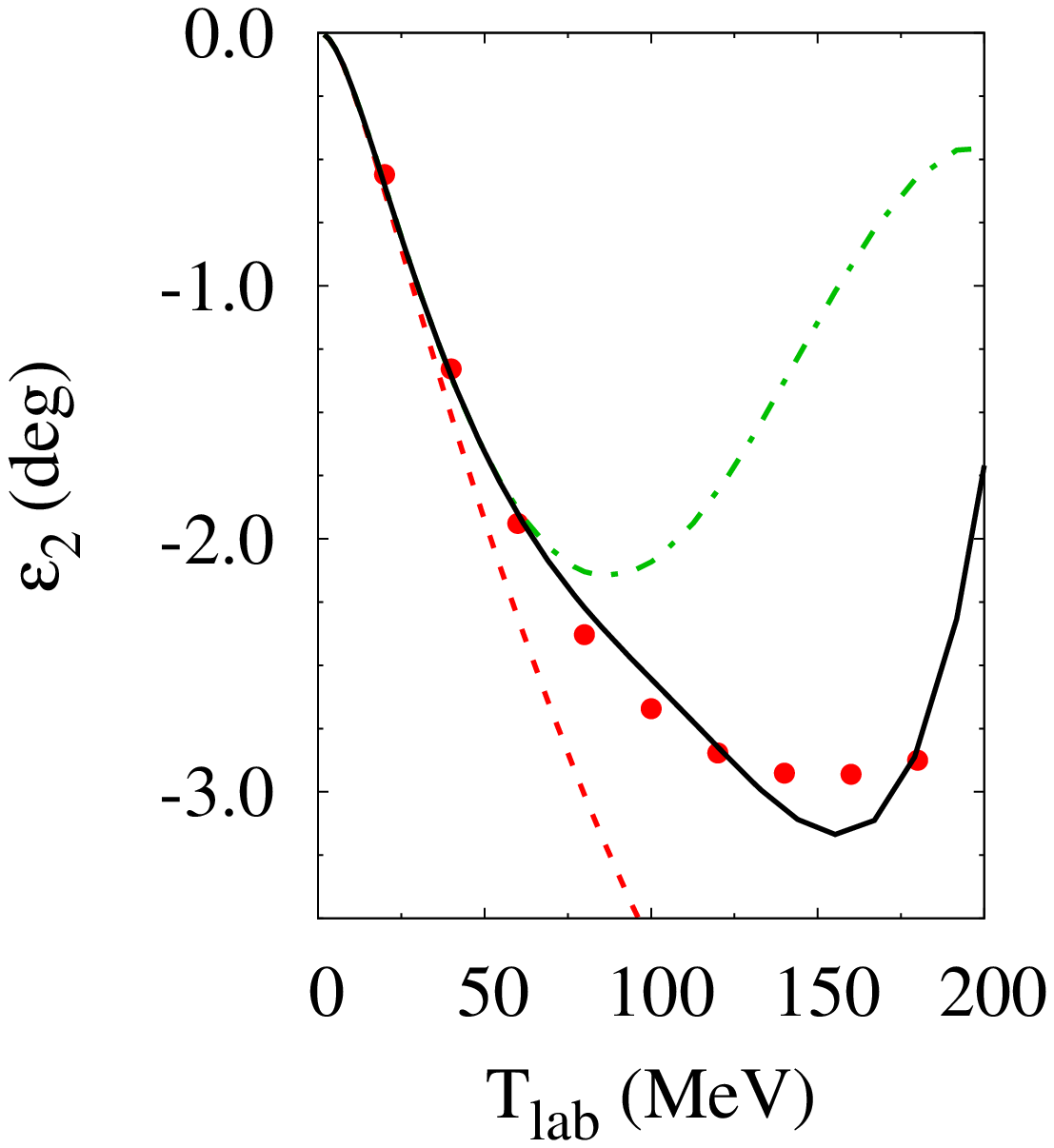}
\caption{(Color online) With the SCTs \eqref{eqn:sub3p2}, the $\cp{3}{2}$, $\cf{3}{2}$ phase shifts and the mixing angle $\epsilon_2$ as functions of $T_\text{lab}$ with $\Lambda = 1.5$ GeV. The values of the counterterms are adjusted to reproduce the PWA values for $\delta_{\cp{3}{2}}$ at $T_\text{lab} = 50$, and $100$ MeV and $\epsilon_2$ at $50$ MeV. The dots are from the Nijmegen PWA. The dashed, dot-dashed and solid lines are $\mathcal{O}(1)$, $\mathcal{O}(Q^2)$, and $\mathcal{O}(Q^3)$, respectively.}
\label{fig:tlab3p2_better}
\end{figure}

To assess the feasibility of improving the fitting quality, we refit the counterterms to the PWA inputs at higher energies: $\delta_\cp{3}{2}$ at $50$ and $100$ MeV, and $\epsilon_2$ at 50 MeV. The updated $\cpf$ EFT phases are plotted in Fig.~\ref{fig:tlab3p2_better}. Although the EFT convergence still breaks down at lower energies than in $\csd$, a good fit to the PWA until $130$ MeV is achieved at $\mathcal{O}(Q^2)$ and $\mathcal{O}(Q^3)$, and it is comparable to the WPC-based calculation with the same TPEs, which is shown in Ref.~\cite{Epelbaum:1998ka-1999dj}. We notice another fitting strategy used in Ref.~\cite{Valderrama:2011mv}, which sacrifices the LO near threshold in order to facilitate better agreements with the PWA at higher orders. This amounts to tuning the LO counterterm to further reduce the attraction of OPE.

Overall, the breakdown scale implied in the numerical results is consistent with our expectation for the delta-less theory, $k \sim \delta \sim 300$ MeV, with the exceptions of $\cp{3}{1}$ and $\cp{3}{2}-^3F_2$. Even for these two channels, one cannot help wondering whether the delta can bring some attraction from $\mathcal{O}(Q^3)$ to $\mathcal{O}(Q^2)$ and improve the convergence of EFT expansion~\cite{daniel-private}.

\section{Discussion and Conclusion\label{sec:conclusion}}

We have studied up to $\mathcal{O}(Q^3)$ the structure of counterterms of chiral $NN$ contact interactions in the triplet channels,  with $S$ and $P$ waves as the examples. The essential guideline we have followed is to promote counterterms over WPC when RG invariance requires it.

We found that the scaling of SCTs are mainly driven by the interplay between TPEs and the LO wave functions. A direct consequence is that $\mathcal{O}(Q)$ by contact interactions alone vanishes. The resulting arrangement of the counterterms in the studied channels are given by Eqs.~\eqref{eqn:3p1ccpresI}, \eqref{eqn:3p0ccpresII}, \eqref{eqn:sub3s1}, and \eqref{eqn:sub3p2}, which can be very nicely summarized by $\mnda$: the SCTs are enhanced by the same amount as the LO counterterm so that the whole tower of counterterms with the same quantum number is shifted uniformly. While this means that WPC remains intact in $\csd$ and $\cp{3}{1}$, it requires an enhancement of $\mathcal{O}(\mhi^2/\mlo^2)$ to all counterterms in $\cp{3}{0}$ and $\cpf$.

It is interesting to compare the chiral $NN$ forces with the three-body system that has only contact interactions. When the two-body $S$-wave scattering length $a_2 \to \infty$, the three-body ``pionless'' theory can be mapped onto a dual two-body theory with $-1/r^2$ long-range force and contact interactions that represent three-body operators in the original system~\cite{efimov}. Reference~\cite{Barford:2004fz} proposed a power counting similar to $\mnda$ for the three-body contact interactions. However, the resemblance between chiral EFT forces and the system investigated in Ref.~\cite{Barford:2004fz} is not perfect because the long-range interactions beyond the leading $-1/r^2$ in the dual two-body system, if any, are resummed nonperturbatively instead of being treated as perturbations. Therefore, it is not clear to us whether the scaling of SCTs obtained in Ref.~\cite{Barford:2004fz} is driven by the LO or subleading long-range interactions.

Our finding that three counterterms are needed in the coupled channels up to $\mathcal{O}(Q^3)$ differs from that of Refs.~\cite{Valderrama:2009ei, Valderrama:2011mv}, which also adopted the perturbative approach on top of the nonperturbative LO, but
concluded instead that six counterterms are necessary for renormalization purpose.
(Although it is speculated in Ref.~\cite{Valderrama:2011mv} that it might be possible to reduce the number of short-range parameters, an alternative is not offered unless the higher-wave component is treated in perturbation theory.) However, that there are only two second-derivative terms---$D_\cs{3}{1}$ and $E_\text{SD}$ in \eqref{eqn:3s1gnrVS}---suggests that one could correlate these six divergent pieces in a model-independent way, which is justified by the numerical evidence of RG invariance shown in Figs.~\ref{fig:fxk-3s1-pres3_012_all} and \ref{fig:fxk-3p2-pres3_012_all}.

Without a dedicated effort to fine tune the fits, our results show a good agreement with the Nijmegen PWA up to $T_\text{lab} \sim 100$ MeV. Regardless of the comparison with the PWA, the relatively large deviation from $\mathcal{O}(Q^2)$ to $\mathcal{O}(Q^3)$ in $\cp{3}{1}$ and $\cp{3}{2}$ encourages one to 
hope that a delta-ful EFT can improve the convergence by including the delta-isobar as explicit degrees of freedom. Aside from the debatable issues of power counting counterterms, the delta-ful nuclear forces have been shown to achieve a more rapid convergence in the two-nucleon~\cite{Valderrama:2008kj_2010fb, Entem:2007jg} and, on a more qualitative level, the three-nucleon~\cite{Pandharipande:2005sx} sectors.

Although there have been many efforts to derive the delta-ful TPEs~\cite{Ordonez:1993-1995, Kaiser:1998wa, Krebs:2007rh}, it appears desirable to update the extraction of low-energy constants in the delta-ful chiral Lagrangian from $\pi N$ scattering through the chiral EFT description around the delta peak~\cite{Long:2009wq, Pascalutsa:2002pi} where the effects of the delta are most prominent. In addition, the formulation of the delta-ful chiral Lagrangian may need to be reexamined in light of the discussion in Ref.~\cite{Long:2010kt}.

\acknowledgments We thank Bira van Kolck and Daniel Phillips for their encouragement in the early stage of this work and thoughtful comments on the manuscript. BwL thanks Jos\'e Goity for useful discussion and for the term ``primordial counterterm.'' CJY thanks Bruce Barrett for valuable support. We are grateful for hospitality to the National Institute for Nuclear Theory (INT) at the University of Washington and the organizers of the INT program ``Simulations and Symmetries: Cold Atoms, QCD, and Few-hadron Systems,'' in which the work was stimulated.
This work is supported by the US Department of Energy under Contracts No.DE-AC05-06OR2317 7 (BwL), No. DE-FG02-04ER41338 (CJY), and the NSF under Grant No. PHYS-0854912 (CJY), and is coauthored by Jefferson Science Associates, LLC under U.S. Department of Energy Contract No. DE-AC05-06OR23177.

\appendix

\section{Perturbative Relation between Phase Shifts and $S$-matrix\label{sec:conversion}}

In distorted-wave expansion, the unitarity of the $S$-matrix no longer rigorously holds. Therefore, the Stapp parametrization~\cite{Stapp:1956mz} of the $S$-matrix in the coupled channel,
\begin{equation}
S = 
\begin{pmatrix}
\cos(2\epsilon) e^{2i\delta_1} & i\sin(2\epsilon) e^{i(\delta_1 + \delta_2)} \\
i\sin(2\epsilon) e^{i(\delta_1 + \delta_2)} & \cos(2\epsilon) e^{2i\delta_2}
\end{pmatrix} \, ,
\label{eqn:stapp}
\end{equation} 
needs to be adjusted, although it is still valid at $\mathcal{O}(1)$ where the LO potential is fully iterated. Here, $\delta_1$ and $\delta_2$ are the phase shifts of the partial wave with $l=j-1$ and $l=j+1$, respectively, and $\epsilon$ is the mixing angle. Suppose that with $S^{(1)}$ vanishing, the $S$-matrix and the phase parameters have the following expansion (labeled by the usual EFT order): 
\begin{align}
S &= S^{(0)} + S^{(2)} + S^{(3)} + \cdots \, , \\
\delta_{1, 2} &= \delta_{1, 2}^{(0)} + \delta_{1, 2}^{(2)} + \delta_{1, 2}^{(3)} + \cdots \, , \\
\epsilon &= \epsilon^{(0)} + \epsilon^{(2)} + \epsilon^{(3)} + \cdots \, .
\end{align}
Expanding both sides of Eq.~\eqref{eqn:stapp}, one finds that
\begin{equation}
S^{(2)} =
\begin{pmatrix}
S^{(2)}_{11} & S^{(2)}_{12} \\
S^{(2)}_{21} & S^{(2)}_{22}
\end{pmatrix}\, ,
\label{eqn:S2}
\end{equation}
with
\begin{equation}
\begin{split}
S^{(2)}_{11} &= e^{2i\delta_1^{(0)}}\left[-2\epsilon^{(2)}\sin(2\epsilon^{(0)}) + i2\delta_1^{(2)}\cos(2\epsilon^{(0)})\right] \, , \\
S^{(2)}_{12} = S^{(2)}_{21} &= i e^{i(\delta_1^{(0)}+\delta_2^{(0)})}\left[2\epsilon^{(2)}\cos(2\epsilon^{(0)}) + i(\delta_1^{(2)}+\delta_2^{(2)})\sin(2\epsilon^{(0)})\right] \, , \\
S^{(2)}_{22} &= e^{2i\delta_2^{(0)}}\left[-2\epsilon^{(2)}\sin(2\epsilon^{(0)}) + i2\delta_2^{(2)}\cos(2\epsilon^{(0)})\right] \, .
\end{split}
\label{eqn:S2abc}
\end{equation}
Replacing the superscript $^{(2)}$ with $^{(3)}$ in Eqs.~\eqref{eqn:S2} and \eqref{eqn:S2abc}, one obtains the relations for $\mathcal{O}(Q^3)$. To convert the $T$-matrix to the $S$-matrix, notice the normalization adopted in the paper:
\begin{equation}
S = 1 - 2i k m_N \left( T^{(0)} + T^{(2)} + \cdots \right) \, .
\end{equation}

\end{document}